\documentclass[preprint,12pt]{elsarticle}

\usepackage{amssymb}
\usepackage{multirow}
\usepackage{siunitx}
\usepackage{booktabs}
\usepackage{adjustbox}

\usepackage[long, nocomma]{optidef}

\journal{Applied Energy}
\usepackage{cleveref}
\crefname{equation}{}{}
\Crefname{equation}{Equation}{Equations} 
\crefname{table}{Table}{Tables}
\crefname{figure}{Fig.}{Fig.}
\Crefname{figure}{Figure}{Figures}
\crefname{section}{Section}{Sections}
\usepackage{xcolor} 
\newcommand{\sh}[1]{{\color{blue}#1}}

\usepackage{nomencl}

\journal{Applied Energy}
\usepackage[ruled, lined, linesnumbered, commentsnumbered, longend]{algorithm2e}
\usepackage{comment}

\begin{document} 
\begin{frontmatter}

\title{Analyzing At-Scale Distribution Grid Response to Extreme Temperatures}
\tnotetext[label1]{This work is sponsored by PNNL, operated by Battelle for the U.S. Department of Energy under Contract DE-AC05-76RL01830.}
\author[inst1]{Sarmad Hanif, Monish Mukherjee, Shiva Poudel, Rohit A Jinsiwale, \\  Min Gyung Yu, Trevor Hardy,  Hayden Reeve}

\affiliation[inst1]{organization={Pacific Northwest National Laboratory},
            addressline={902 Battelle Boulevard}, 
            city={Richland},
            postcode={99352}, 
            state={WA},
            country={USA}}

\begin{abstract}
\noindent Threats against power grids continue to increase, as extreme weather conditions and natural disasters (extreme events) become more frequent. Hence, there is a need for the simulation and modeling of power grids to reflect realistic conditions during extreme events conditions, especially distribution systems. This paper presents a modeling and simulation platform for electric distribution grids which can estimate overall power demand during extreme weather conditions. The presented platform's efficacy is shown by demonstrating estimation of electrical demand for 1) Electricity Reliability Council of Texas (ERCOT) during winter storm Uri in 2021, and 2) alternative hypothetical scenarios of integrating Distributed Energy Resources (DERs), weatherization, and load electrification. In comparing to the actual demand served by ERCOT during the winter storm Uri of 2021, the proposed platform estimates approximately 34 GW of peak capacity deficit\footnote{These numbers are consistent with state-of-the-art prediction results published in the literature \cite{gruber2022profitability}}. For the case of the future electrification of heating loads, peak capacity of 78 GW (124\% increase) is estimated, which would be reduced to 47 GW (38\% increase) with the adoption of efficient heating appliances and improved thermal insulation. Integrating distributed solar PV and storage into the grid causes improvement in the local energy utilization and hence reduces the potential unmet energy by 31\% and 40\%,
respectively.
\end{abstract}

\end{frontmatter}

\section{Introduction}\label{sec:introduction}
Extreme events are becoming more and more frequent, to the level that they may not be ignored while planning and operating future power systems \cite{wang2015research}. Unsurprisingly, planning and operating the power grid to account for these extreme conditions is becoming one of the most critical challenges for power system's planning and operation entities (\textit{e.g.}, system operators and utilities) \cite{brown2013preparing, reuters2021, federal2011report}. Similarly, the need for methods and tools to help plan and operate power systems for extreme weather events is also of immense need \cite{bennett2021extending}. As the fundamental planning variable for the power grid is the electricity demand, estimating and predicting a region's electricity demand during an extreme event may prove crucial for deploying resources to combat against predicted extreme events. Moreover, the adversity of an extreme event is usually characterized by the amount of unmet load, before, during and after the event and consequently how many customers were left without access to power, and for how long. Hence, the ability to gauge the increase in system demand due to an extreme event scenario is of great importance to grid planners and analysts. This paper presents the results of a modeling and simulation platform which captures distribution grid demand during an extreme weather event. Utilizing the platform, the paper then explores distribution grid conditions, such as losses and grid violations, during such an event and system net demand under hypothetical DER and retrofit scenarios. We chose to focus on the distribution grid demand modeling for two reasons. First, distribution grids face the greatest level of disruption during extreme events~\cite{MUKHERJEE2018283, kwasinski}. Second, distribution grids are located closest to the consumers, which allows more detailed modeling of consumer actions and strategies to better represent the overall demand dynamics. Although there can be many types of extreme conditions, (\textit{e.g.} hurricanes, super-storms etc.), this paper focuses on  extreme temperatures. This is motivated by the fact that even though extreme temperatures may not be considered as a natural disaster, they may end up causing the same level catastrophic disruptions (\textit{e.g.}, see the impacts of a recent events of winter storm Uri in Texas \cite{BUSBY2021102106}\footnote{Heat waves are potentially also fall into this category, but historically have been dealt better by utilities, e.g. see California responses~\cite{climateimpact, stone2021climate}).}.

In the literature there exists two general approaches to estimating large, regional electricity demand due to extreme temperature. First are the top-down models which may consist of deploying regression expressions of demand against historical temperatures~\cite{gruber2022profitability}, machine-learning methods to predict outages~\cite{demandmodeling, machinelearningrozhin}, and classical short-term forecasting techniques~\cite{shortermdemandforecast}. For extracting large-scale grid response, these models are adequate, as the diversity of the large population averages out the individual, sometimes atypical, demands. However, these models do not represent detailed physical knowledge of consumers and hence are unable to be predict demand for conditions for which limited to no data exists. As a result, such models may not be suited to include impacts of resilience-mitigation actions, such as demand response techniques, and local generation adoption. The second approach, modeling from the bottom-up, uses physics-based demand modeling to represent the physical behavior of electrical devices. For example, modeling temperature dynamics inside a house to predict the operational state of an HVAC system and it's corresponding power consumption. Authors in~\cite{stone2021climate, buildings11030096} demonstrated how energy requirements for buildings can be estimated using a physics-based building energy simulation tool. Authors in \cite{stone2021climate} focused on heat waves, whereas \cite{buildings11030096} utilized the energy models of commercial building to develop occupant's resilience metrics. Note that these physics-based modeling environments may also be used to generate data for regression-based/machine-learning models and it may end up providing further insights on variables governing electricity demand during extreme events.

The above outlined approaches for estimating power demand during extreme events do not consider power-system-specific interactions. These interactions are important for not only representing the impact of grid mechanisms (\textit{e.g.} fault isolation) during a resilience event, but also to include grid physics (\textit{e.g.} losses) which may impact the overall electric power demand. Transmission grid models and mechanisms have been deployed in the literature, such as in~\cite{syntheticgriddatatexas}. These models have been developed, following the success of large-scale open-source synthetic grid models~\cite{yixinplatform, bornsheuer2019large} and power flow modeling tools~\cite{Zimmerman:2011um}. The choice of a transmission grid model to determine grid response may be motivated by managing the complexity of the model, as extreme events impact a large fraction of population. However, as discussed earlier, distribution grid representation is crucial for estimating demand, as most of the outages happen at the distribution grid level and directly impact consumers connected to it. Moreover, distribution grids are inherently very different than transmission grids. For example, in distribution grids, conditions such as voltage drops due to higher loading level and resultant losses occur more regularly as compared to transmission grids. Similarly, there may exist very different timing and impacts of a utility's resilience measures in distribution grids which may impact the overall unmet load during a resilience event. To this end, a large co-simulation platform is proposed to capture the interaction of both the transmission and distribution systems~\cite{TESP, staid2021north}. However, existing platforms' distribution grid models are not yet tested in terms of 1) benchmarking it against a known extreme temperature event, 2) investigating alternative future DERs and load electrification scenarios and observing their impact on key event metrics. 

In this paper, we address the above-mentioned need for a reliable at-scale distribution grid response estimation on two fronts. First, we present a modeling and simulation platform that supports physics-based load models in a distribution grid simulation environment to demonstrate large-scale system response to extreme temperatures. Second, to support analytical capabilities of the presented platform, alternative scenarios of future DERs, building weatherization, and load electrification are analyzed in terms of their impact on distribution grid demand estimation during an extreme weather event. The analysis performed in this paper consists of reproducing Texas conditions during winter storm Uri of 2021 and is compared against published data for predicted loads in \cite{gruber2022profitability} and actual outage data from Electricity Reliability Council of Texas (ERCOT) in \cite{ercot_hist}. To limit the scope, this paper focuses on physical power grid interactions (\textit{e.g.} reactive power flows, losses, and violations) impacting the overall power demand and does not consider conventional (\textit{e.g.} switch operations, feeder isolation etc.) and contemporary (\textit{e.g.} utilizing demand flexibility) measures for coping with resilience. Hence, to allow for future work aiming to investigate conventional and emerging resiliency strategies (such as demand flexibility) the presented modeling and simulation platform capture the dynamics and control of the distribution system and DERs.

The rest of the paper is organized as follows. \Cref{sec:modelingframework} explains the modeling and simulation platform used in this paper. \Cref{sec:scen_analysis} demonstrates the scenarios analyzed in the paper. \Cref{sec:case-study} presents results and \cref{sec:conclusion} concludes the paper and provides plausible future works directions.
\begin{figure}[ht]
    \centering
    \includegraphics[width=0.95\textwidth]{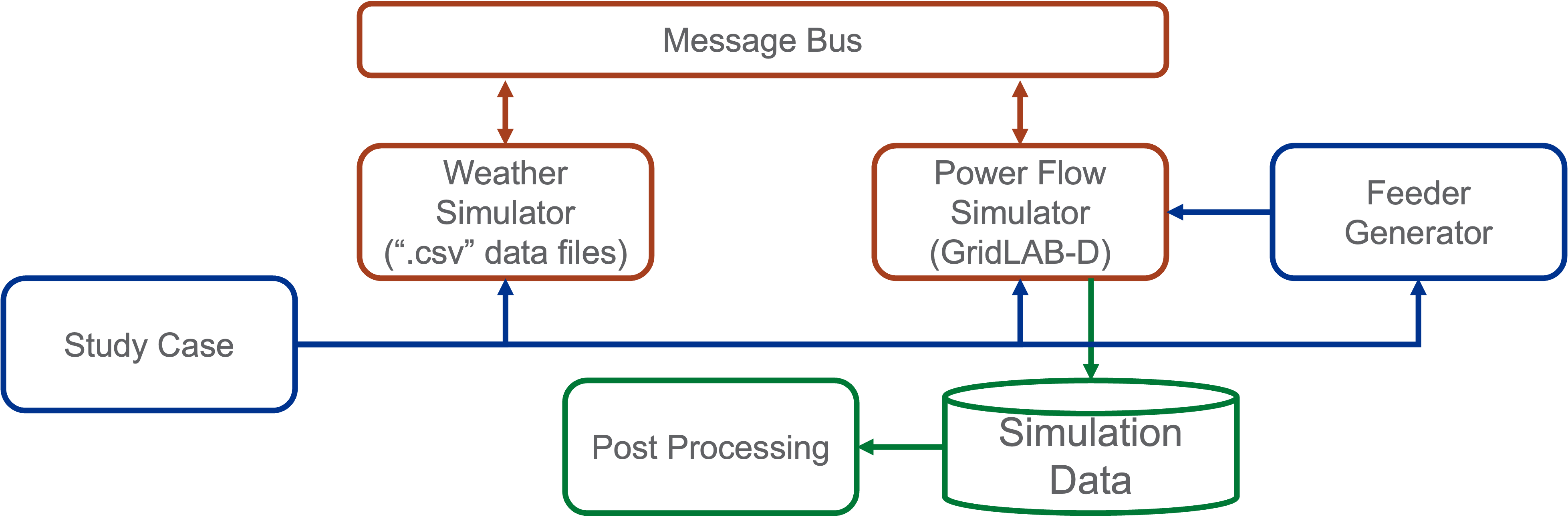}
    \caption{Modeling and simulation platform of this paper.}
     \label{fig:modeling_simulation_platform}
\end{figure}
\section{Modeling and Simulation Platform}\label{sec:modelingframework}
This paper's presented modeling and simulation platform is a customized version of the Transactive Energy Simulation Platform (TESP)~\cite{tesp_github}, which is a co-simulation tool, allowing multiple state-of-the-art simulation tools to exchange information with each other and obtain a high-fidelity power system simulation. \Cref{fig:modeling_simulation_platform} gives an overview of the components from (TESP) that were utilized in this paper. The platform consists of two simulators: 1) a weather simulator and 2) a power flow simulator. A message bus is used to exchange information between these simulators. The overview of the component of the platform is provided next. Refer to \cref{sec:appendix} for overview on the key modeling aspects of the platform and \cite{reeve2022dso+} for the detailed information on population calibration for large-scale simulation study.
\subsection{Power Flow Simulator}
\paragraph{Load Modeling}\label{sec:powerflowsimulator}
GridLAB-D\textsuperscript{TM} is central to this paper's modeling and simulation platform, as it provides the capability to obtain weather-dependent load profiles of distribution grids power demand. This is enabled by GridLAB-D's  modeling of temperature-dependent thermostatic loads, such as water heaters and the heating, ventilation, and air-conditioning (HVAC) systems inside a building's thermal envelope with modeled thermodynamics. In general, GridLAB-D uses a mixture of ZIP (constant impedance, current, and power) loads, plug loads, and thermostatic loads to represent the total load for a single house. As the HVAC load is one of the largest loads and is highly sensitive to outdoor temperatures, we present a brief overview of its modeling procedure in \cref{sec:app_etp}. The implementation of the latest GridLAB-D residential load models can be found in \cite{gridlabd_github_residential} and the accuracy of the model to represent residential load dynamics has been provided in \cite{carolyn}. For this paper, GridLAB-D is used to first calibrate/fine-tune distribution grid load against historical data and then used to extract grid response under various hypothetical scenarios. The following characteristics of GridLAB-D allows to perform such analysis. 
\begin{itemize}
    \item There are three heating technologies modeled in GridLAB-D: gas heater, heat pump, and resistance heating. As heat pumps are more efficient than resistance heating, the percentage of heating equipment ownership can be adjusted and its impact on the overall distribution grid load can be measured. 
    \item GridLAB-D utilizes the Equivalent Thermal Parameter (ETP) model to estimate the HVAC load for the house. The ETP model parameterizes house insulation using ``R’’ values, which can be adjusted to represent a high or a low insulation level of the house. These values can be adjusted to obtain an appropriate response of house insulation on the distribution grid load.
\end{itemize}

\paragraph{Feeder Modeling}\label{sec:feedermodeling} In order to set up the case-study to perform the desired simulation and grid response, prototypical feeders \cite{Schneider2008taxonomy} are populated with GridLAB-D house models. Appropriate house models, based on the combination of water heater, HVAC, lighting load, and plug loads with their activity schedules (e.g., water draw, internal mass due to occupancy, etc.) were specified. The rating of the loads and the activity schedules are assigned based on the statistical distribution of the housing population of the region to be modeled. Based on the selected house model, the service transformers, fuses, and circuit breakers are also sized to allow the distribution feeder to host the additional load. 
Relevant to the resilience analysis, the typical feeder modeling procedure described above was modified. This was done because during the preliminary analysis of power flow conditions of the feeders, it was observed that the ratings of the distribution system components (lines, transformers) of the prototypical feeders were not adequate enough to generate the peak load which was observed during the Winter storm Uri. Though for this level of abnormally high load, due to tripping of certain protection equipment, it may be a realistic response of the distribution grid, however, 1) it does not help for the goal of this paper to provide an accurate estimation of the required level of infrastructure upgrade to avoid such an extreme condition, 2) as well as does not allow to estimate the maximum aggregated response of the grid to the extreme event. As these evaluations are the goal of this paper, we oversize transformers and lines to allow for hosting such large loads in the distribution grid.

\paragraph{Weather Simulator} As shown in \cref{fig:modeling_simulation_platform}, the required external weather information for simulating distribution grid response is coordinated by the message bus through the weather simulator. For this work, time-indexed files in a ``.csv’’ format are used by the weather simulator which provides information such as temperature, humidity, solar insolation, pressure, and wind speed, to be utilized as input for the power flow simulator. In \cref{sec:res_case}, an example of how weather data was collected to generate the required extreme weather information, is to be included in the presented platform.

\begin{figure*}[!tb]
    \centering
\includegraphics[width=0.95\textwidth]{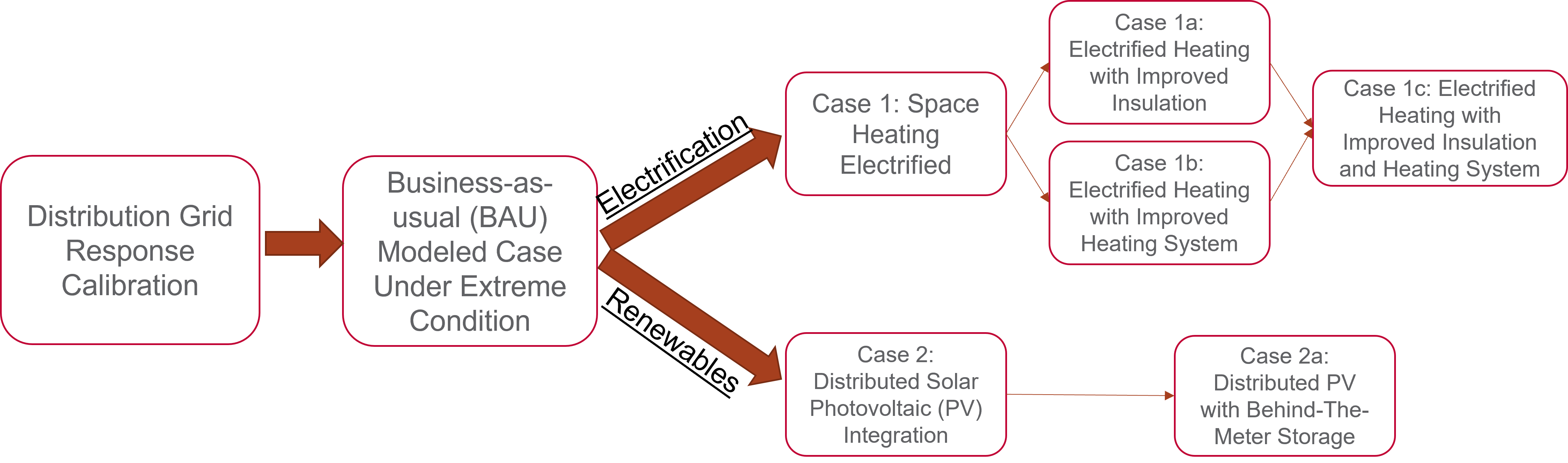}
    \caption{Modeled cases of the paper.}
     \label{fig:overview_picture}
\end{figure*}
\section{Scenario Analysis}\label{sec:scen_analysis}
\Cref{fig:overview_picture} shows the overall scenarios presented in this paper. First, a simulation configuration for the region of interest is fine-tuned to obtain a calibrated response model of the respective grid area of the interested region. Following the modeling and simulation additions explained in \cref{sec:modelingframework} to capture the impact of extreme weather event, the calibrated simulation model is simulated using the extreme weather. This represents the business-as-usual (BAU) distribution grid model's response to extreme weather. Using this model for representing system demand for extreme conditions, we present two future load pathways, 1) an electrification pathway and 2) a higher renewable DER deployment pathway. We demonstrate the impacts of these pathways on the overall system demand during the extreme weather condition. The modeled region for all scenarios is taken as the ERCOT region, with the extreme condition taken as winter storm Uri of 2021.

\paragraph{Modeling Assumptions}

As the focus of this paper is on demonstrating the capability of the  modeling and simulation platform to reproduce, predict and estimate the impact of extreme events under different grid configurations, the following assumptions are made:
\begin{enumerate}
    \item To obtain a counter-factual baseline response of the grids if they had not failed, it is assumed that consumer devices and grid infrastructure continues to operate during an extreme event. That is, no damage to utilities and consumers is assumed and consequently, no loss of load is assumed. 
    \item For Solar PV and behind-the-meter battery, the efficiency do not reduce due to snow. However, solar PV profiles are generated using temperature and irradiation data (as described in \cref{sec:res_case}), which is represented by Texas' actual recorded temperature, demonstrating reduction in available PV power production during the winterstorm Uri.
    \item For HVAC systems, no external damages due to extreme weather is assumed. However, as a direct consequence of modeling the Coefficient of Performance (COP) as a function of operating condition, HVAC response does reflect performance degradation as a result of operating in extreme winter temperatures
    \item No advanced control algorithms for coordinating the response of the DERs are explored and only the local controllers built in the power flow simulator (GridLAB-D) are modeled. These controllers include, the thermostat controller of HVAC and water heater. The unity power factor controller for Solar PV and battery. The legacy voltage controllers (discrete step-size) for capacitor banks and transformer regulators. While this help in obtaining a realistic distribution grid response as the commonly available local controls in the grid, it also equips the analysis to be compared against future external control development to improve grid response to extreme weather, e.g. controls to extract demand-side flexibility.
    \item As shifting demand (exercising demand-side flexibility) is out of the scope of this paper and is recognized as a key future work, we assume that consumers pre-defined thermostat settings remain same during the extreme weather event. Adjusting these controls to adapt to extreme temperatures is recognized as a key future work.
    \item The control of the behind-the-meter storage is assumed to be done by a utility and it is assumed that it stays invariant during an extreme event. It is assumed that storage owners are instructed by the utility to charge/discharge during day/night-time to take advantage of the excess/unavailable solar power and there exists enough incentive for the consumers to participate in this activity. It is assumed that the utility provides signals for charge/discharge to consumers with enough diversification to avoid creating a new peak of charge/discharge.
\end{enumerate}

\begin{figure}[h]
    \centering
    \includegraphics[width=0.95\textwidth]{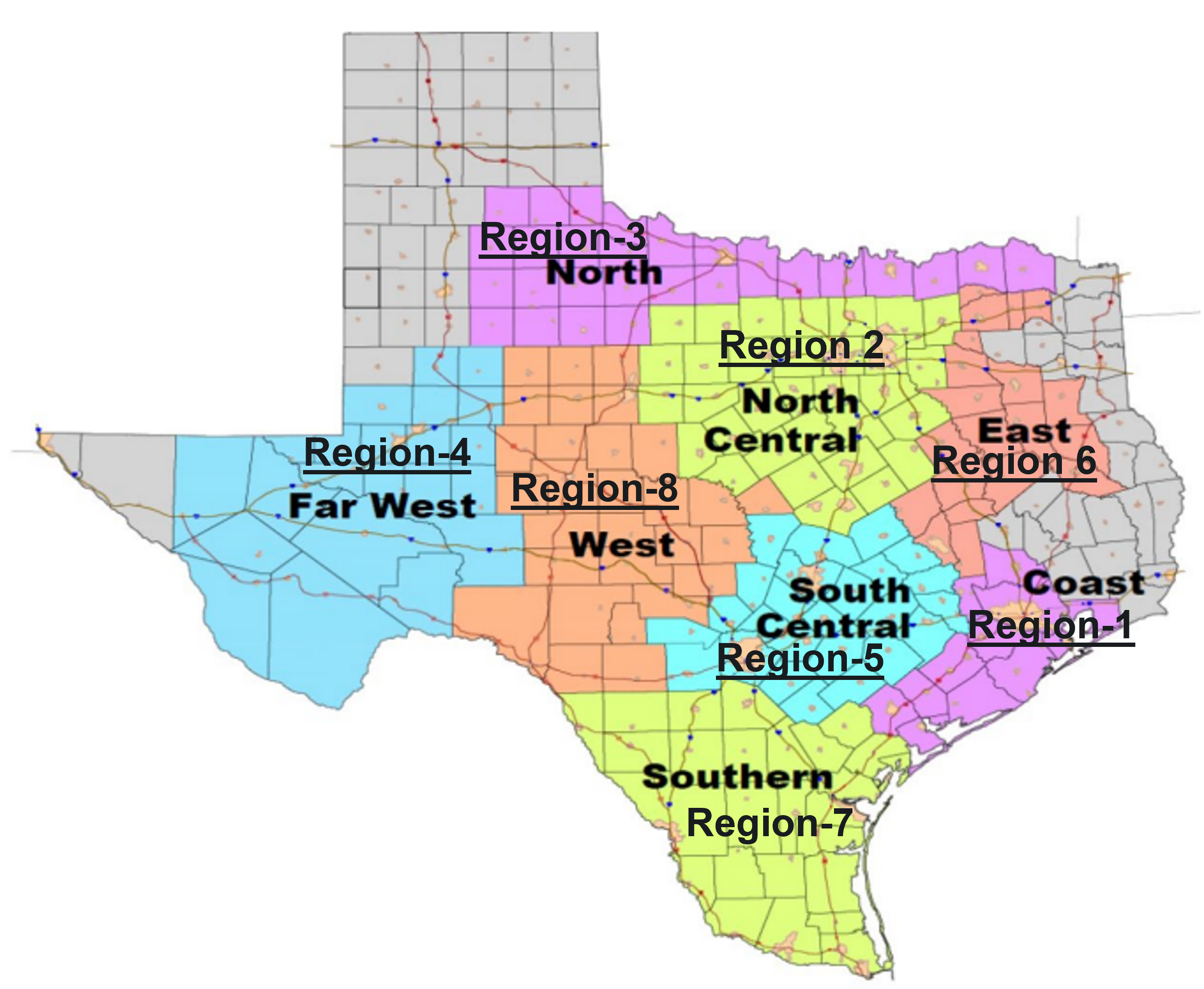}
    \caption{8-Node Model of ERCOT region}
     \label{fig:8node}
\end{figure}
\subsection{Distribution Grid Response Calibration} \label{sec:bau}
This case is modeled to calibrate the response of the distribution grid, by comparing it with the recorded historical data. \Cref{fig:8node} shows the modeled region of this paper using 8 representative regions in Texas. The distribution grid modeling of these regions follows the procedure outlined in~\cite{reeve2022dso+}, and its brief overview is given in \cref{sec:app_model}.

\subsection{Distribution Grid Business-As-Usual (BAU) Response Under Extreme Event Modeling}\label{sec:res_case}
\begin{figure}[h]
    \centering
    \includegraphics[width=0.90\textwidth]{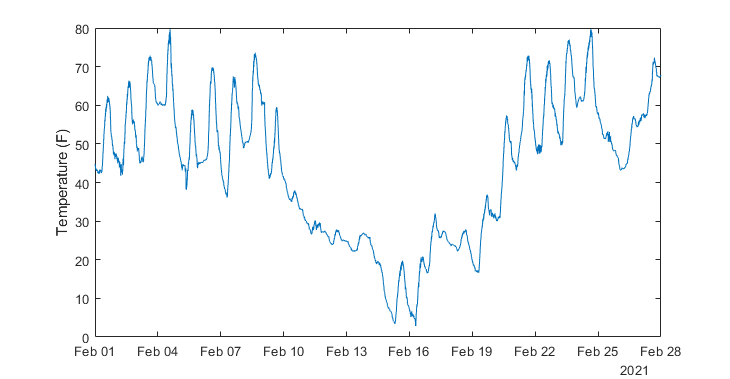}
    \caption{February 2021 Temperature in Dallas,TX}
     \label{fig:Dallastempprofile}
\end{figure}

This case demonstrates ``what would have happened if there was no load-shed in the region during an extreme event -- given the existing grid infrastructure and load composition?'' The case uses same customer population as in case 1. As shown in \cref{fig:8node}, each ERCOT region is assigned a specific weather region~\cite{ercot_hist}. This information was used to collect weather data for the 2021 Texas Vortex Freeze condition. Data for 5-minute resolutions temperature, humidity, and wind speed data was sourced from NOAA archives~\cite{NOAAdata}, and average atmospheric pressure data from \cite{weatherunderground} and was assigned to the geographically closest region.
Fig. \ref{fig:Dallastempprofile} shows the temperature profile for the region modeled after the Dallas area during the extremely low-temperature event observed in Texas in February 2021.

\subsection{Case 1 -- Electrification of Space Heating}
This case models the cases when all the population uses electric heating systems (heat pump or resistance) for electrification. \Cref{fig:heating_system_comparison} shows approximately 36\% of end-users in the ERCOT region uses gas heating, which is distributed evenly between heat-pump and resistance heating. 
\subsubsection{Case 1a -- Space Heating Electrification with Improved Insulation}
\begin{figure}[h]
    \centering
    \includegraphics[width=0.95\textwidth]{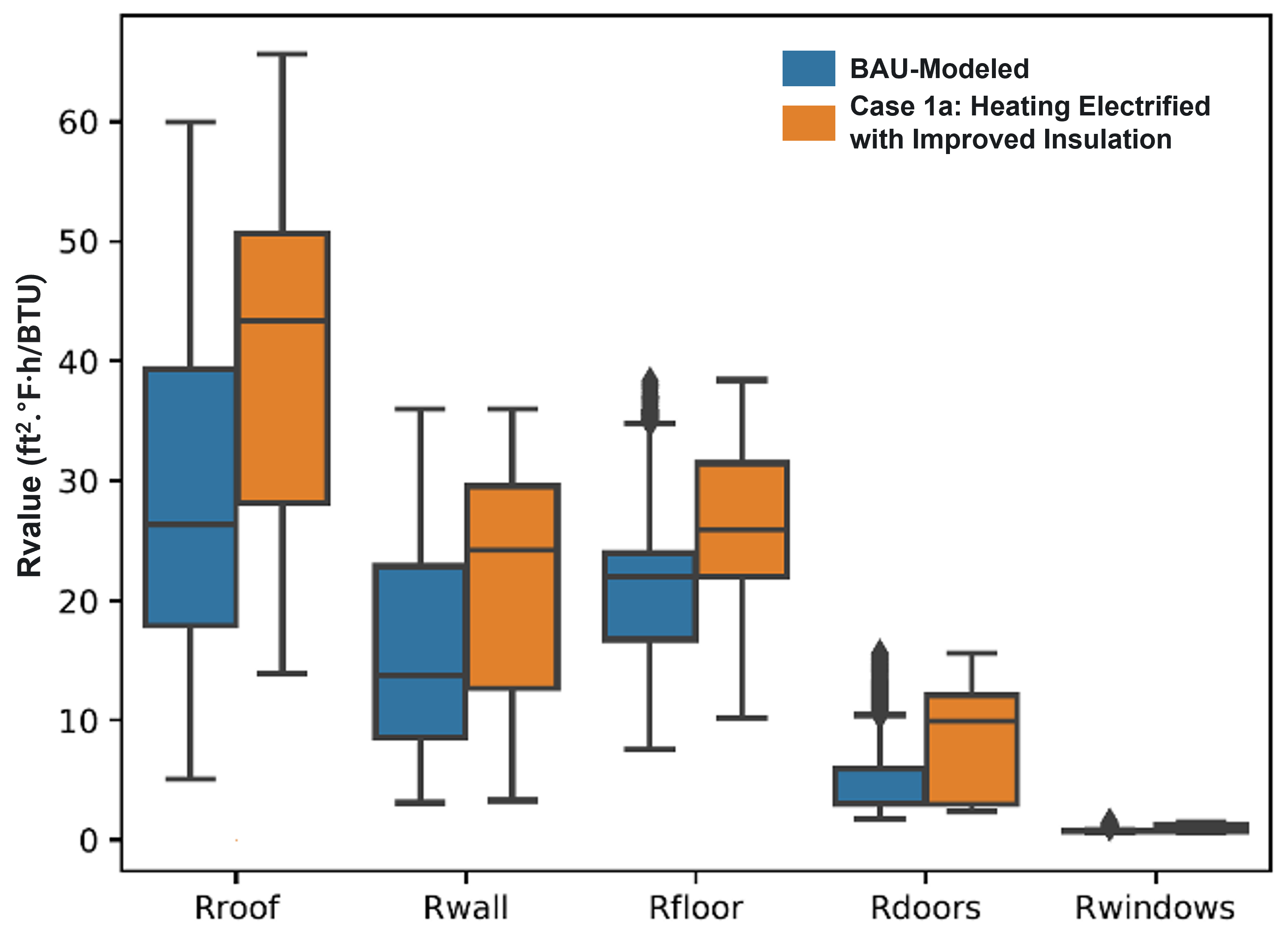}
    \caption{R-values comparison between standard population and high-insulation population}
     \label{fig:Rvalue_comparison}
\end{figure}

This sub-case models improved insulation of building stock in addition to the electrification of space heating. This is done by assuming that buildings meet code requirements consistent with construction after the year 2000 and hence adopt higher insulation levels in their thermal envelope. As a result, the chosen R-values of this scenario are shown in \cref{fig:Rvalue_comparison}, where it can be seen that they on average increase by approximately 22.9 to 63.9 \% for this highly-insulated case. Therefore, it is expected that this scenario case will have lower heating demand during extreme weather. Even though thermal insulation levels are improved, the heating system technology ratio (heat pump versus resistance heating) distribution is kept the same as the standard electrification case 1.
\subsubsection{Case 1b -- Space Heating Electrification with Improved Heating Technology}
\begin{figure}[h]
    \centering
    \includegraphics[width=0.95\textwidth]{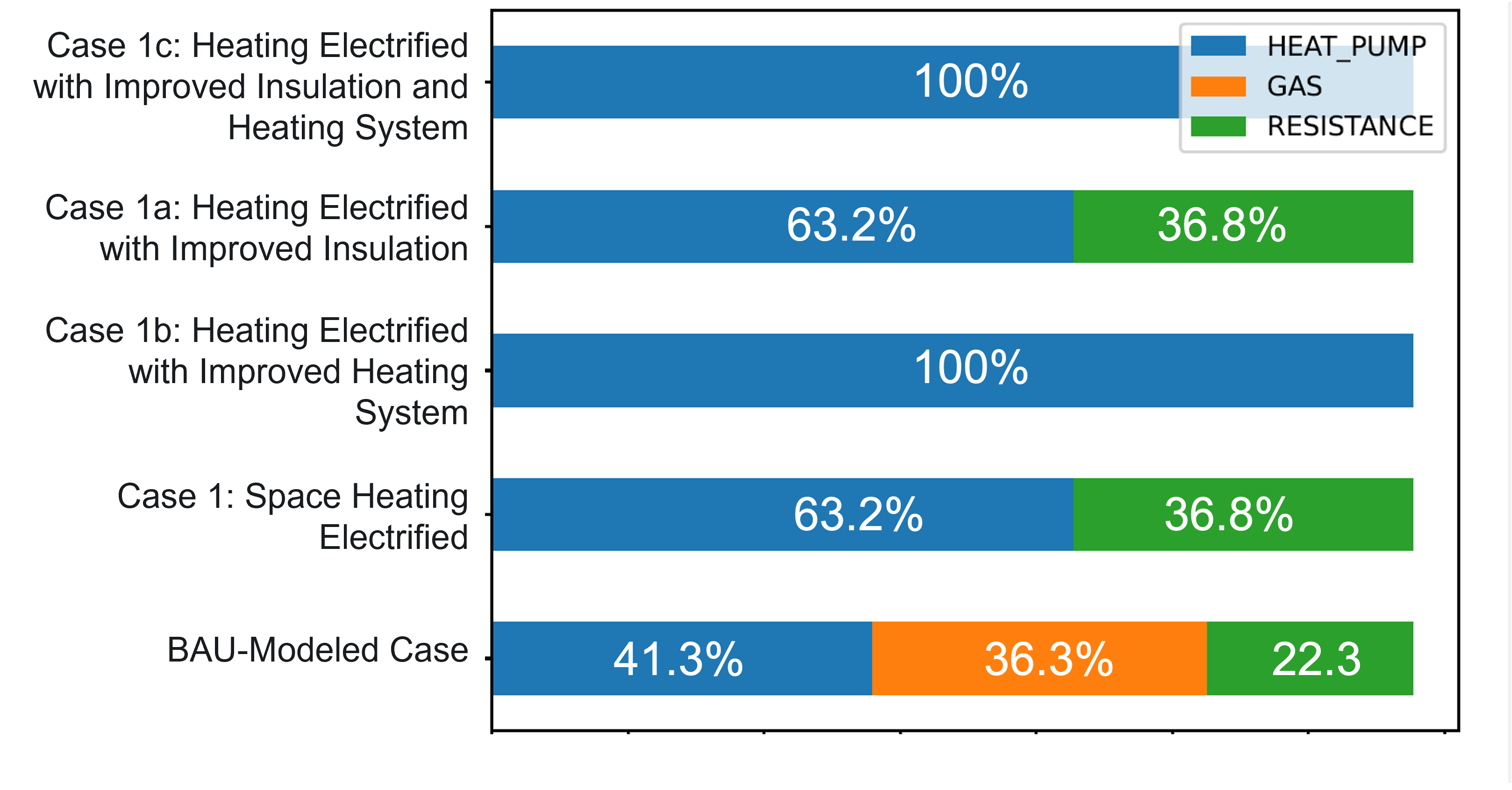}
    \caption{The fraction of heating system comparison in cases}
     \label{fig:heating_system_comparison}
\end{figure}

This scenario models the customer population without resistance heating, i.e., all customers are equipped with heat pump systems for heating to maximize the system efficiency. See \cref{fig:heating_system_comparison} for comparison of heating technologies installed under different scenarios. To not have a bias in the intra-technology efficiency improvement, the efficiency of a heat pump (COP) is still populated using the statistical distribution of the business-as-usual scenario.

\subsubsection{Case 1c -- Space Heating Electrification with Improved Heating Technology and Insulation}
This scenario combines the improved thermal insulation and heating technology scenarios to demonstrate the grid response under efficient electrification directives. That is, all the population uses heat pumps for heating as well as retrofitted high-insulated buildings.

\subsection{Case 2 -- Integration of Distributed Solar PV}
This scenario augments the resilience scenario with distributed solar PV. To model the distributed solar PV, 40\% of houses are assumed to have rooftop solar PV panels. To demonstrate the geographic diversity, randomly generated azimuth angle, tilt angle, and geographic location was deployed in PySAM's PVWatts calculator \cite{pvwatts} to produce a 5-minute solar profile for each region. Using the statistics on distributed PV penetration rate and total customers, solar profiles are scaled to be inputted in GridLAB-D. The detail of distributed modeling is provided in \cite{reeve2022dso+}. All distributed solar PV is assumed to be operating under a unity power factor.

\subsubsection{Case 2a -- Integration of Distributed Solar PV with Behind-the-Meter Storage} 
This scenario includes behind-the-meter distributed storage across 50\% of houses in the grid. It is assumed that no house gets then more than 1 battery. Each battery is modeled as a direct-current electro-chemical device with a specified charge/discharge efficiency, inverter efficiency, power rating, and energy rating. The battery population follows the distribution of a mean 13.5/5 kWh/kW rating with a +/- 20\% range. All batteries are modeled with discharge efficiency of 96\% and inverter efficiency of 98\%.

As the goal of this scenario is to complement distributed solar PV in the distribution grid, the charge and discharge signals for batteries are generated based on observing the solar PV profiles. That is, the charging signal is sent during the day times and discharged during the evening/night-time. This process is automated using another feature of GridLAB-D, where a randomized schedule for each object can be generated by defining the base schedule and individual object's skew parameter to represent shift (delay/advance) from the base schedule. Following this, we generate each battery's charge/discharge schedule using a base charge/discharge schedule and a random skew parameter (+/- 2 hours).

\section{Results}\label{sec:case-study}
\subsection{Distribution Grid Model Response Calibration}
\begin{figure}[!h]
    \centering
    \includegraphics[width=0.95\textwidth]{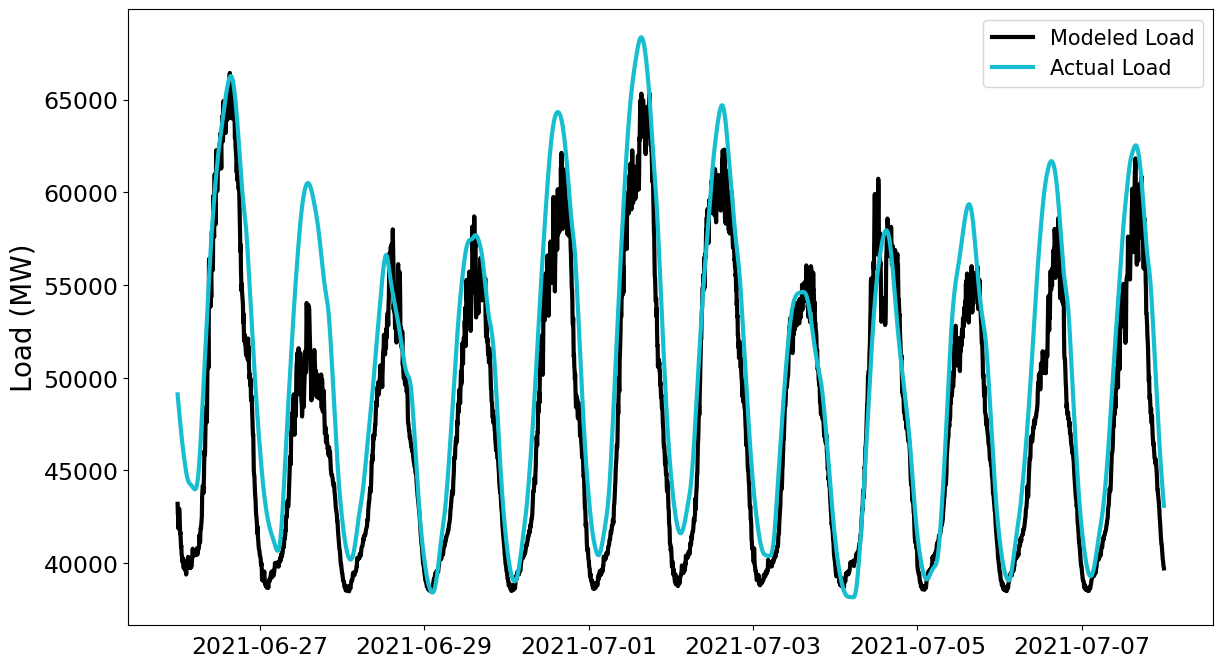}
    \caption{Simulated versus recorded load comparison for summer season}
     \label{fig:time-series-bau-summer}
\end{figure}

\begin{figure}[!h]
    \centering
    \includegraphics[width=0.95\textwidth]{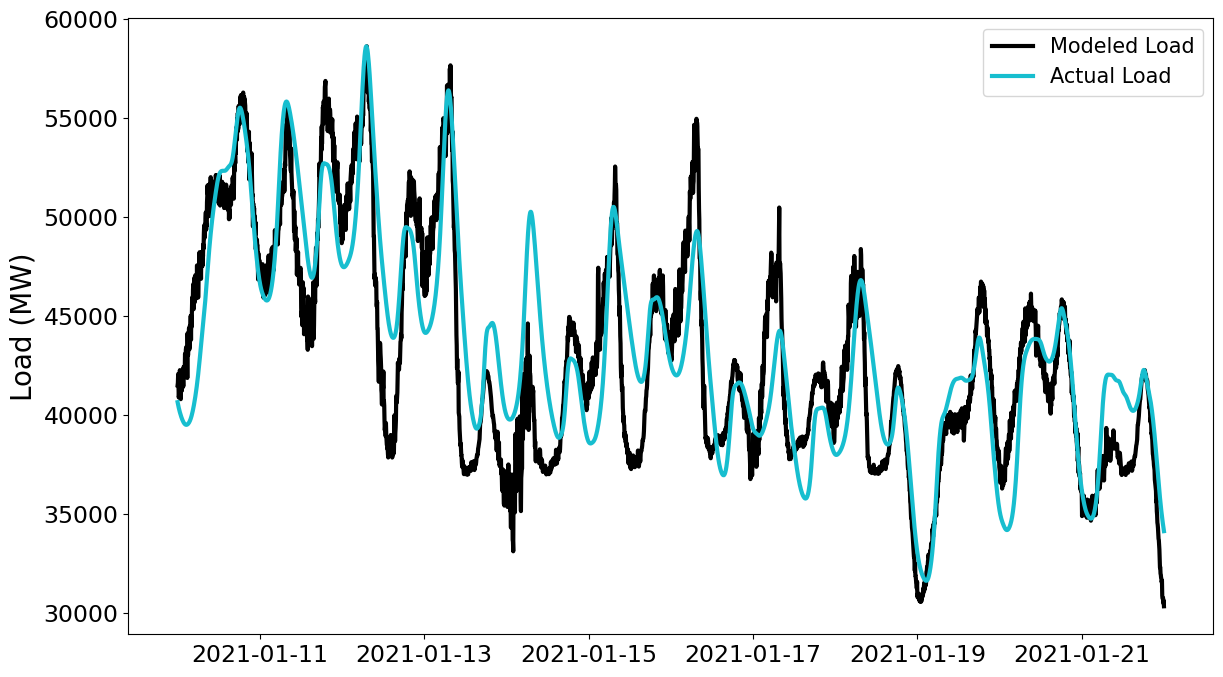}
    \caption{Simulated versus recorded load comparison for winter season}
     \label{fig:time-series-bau-winter}
\end{figure}

The historical data for comparison is taken from \cite{ercot_hist}.
To demonstrate that the presented platform is versatile and can represent diverse grid conditions, first, we present a comparison against business-as-usual conditions. \Cref{fig:time-series-bau-summer} and \cref{fig:time-series-bau-winter} shows comparison for a typical 10 days of summer and winter load. As a reference of how the simulated load changes with the temperature, average of modeled 
8 weather regions temperature profiles is also plotted in \cref{fig:time-series-bau-summer} and \cref{fig:time-series-bau-winter}\footnote{Note that in the actual simulation, each region gets to have its own weather simulation.}. For more insights on the potential validity of simulation modeling and platform of this paper, interested readers are referred to \cite{reeve2022dso+5}, where a much more detailed population calibration and fine tuning for the ERCOT region for for the year 2016 was performed.

Both comparisons demonstrate that simulated demand matches well with the historical data (within approx. 10\% and 20\% error of instantaneous peak and valley loading level prediction for summer and winter population, respectively). However, the simulation does perform better for the summer-time period.
\begin{figure}[!h]
    \centering
    \includegraphics[width=0.95\textwidth]{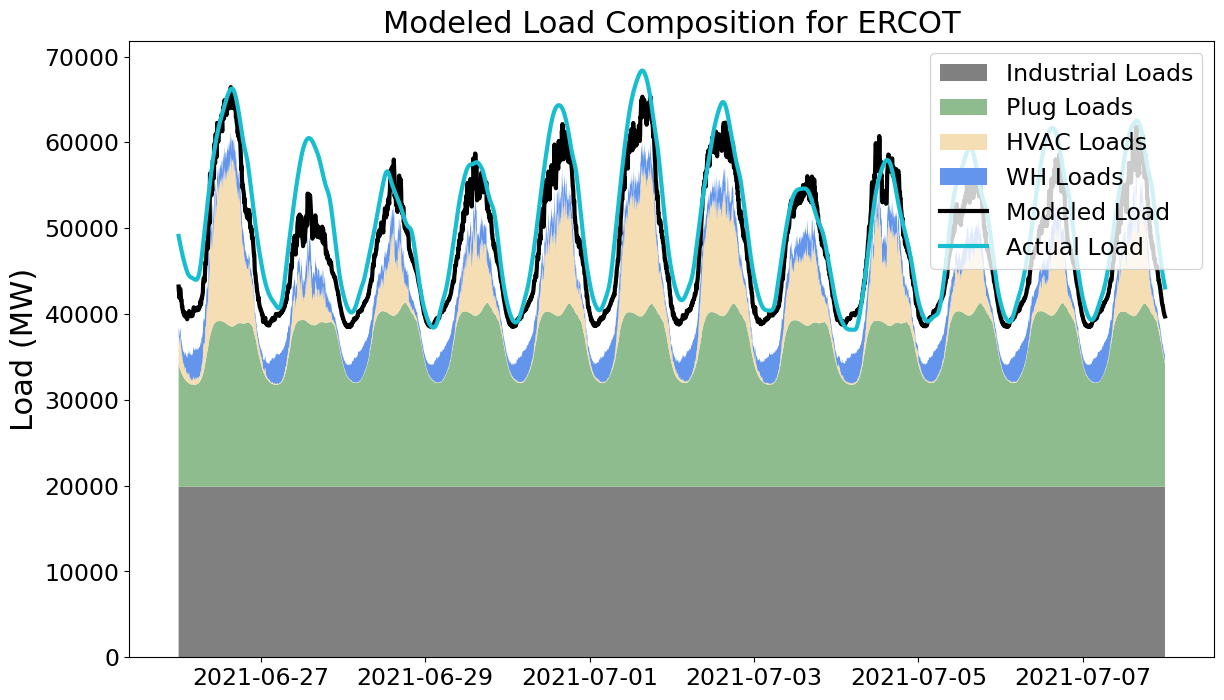}
    \caption{ERCOT load decomposition for summer time period.}
     \label{fig:dso2-summer}
\end{figure}
\begin{figure}[!h]
    \centering
    \includegraphics[width=0.95\textwidth]{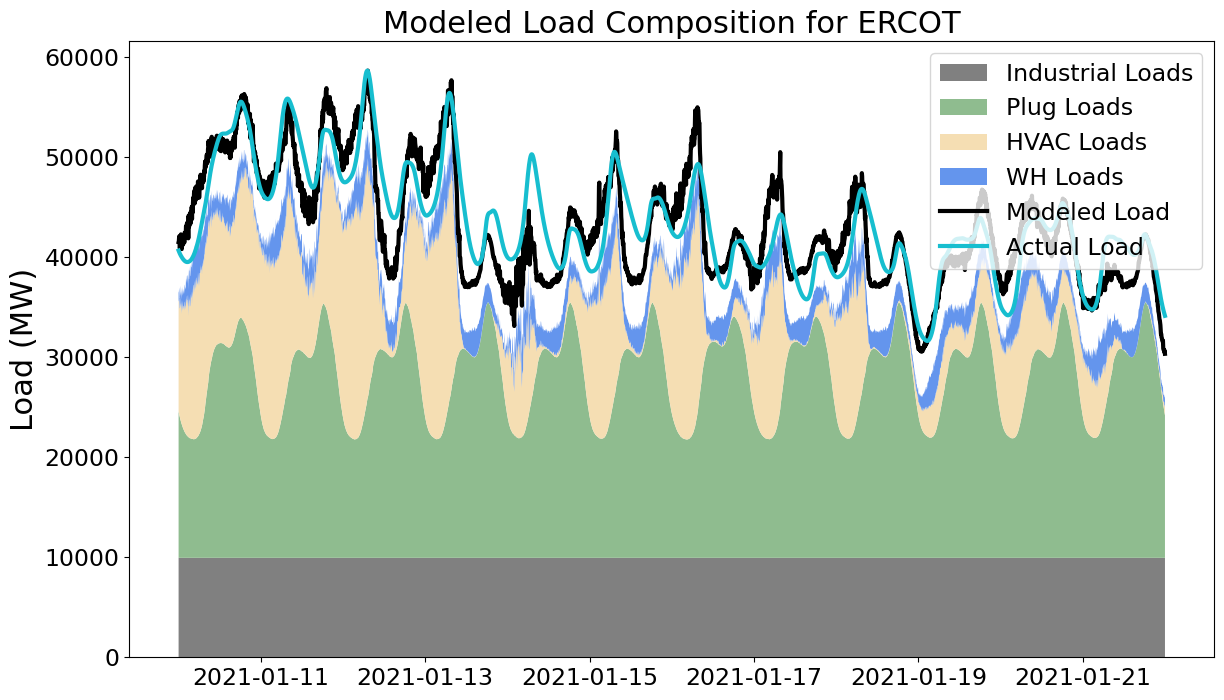}
    \caption{ERCOT load decomposition for winter time period.}
     \label{fig:dso2-winter}
\end{figure}
For both seasons, ERCOT load composition in terms of its modeled end-uses is shown in \cref{fig:dso2-summer} and \cref{fig:dso2-winter}. The simulated aggregated load profile for the ERCOT region is also compared with the recorded load. Due to the modeling and simulation platform's capability to model weather sensitive-load, it can be seen that due to cooling and heating needs, HVAC load increases during day-time in summer and during night-time in winter, respectively. Note that, a small offset between the sum of all end-use load and ``Modeled Load'' in \cref{fig:dso2-winter} is due to the distribution grid losses. As mentioned in the motivation section of this paper, this capability has been made possible due to the deployment of power flow simulator in the simulation and modeling platform.
\begin{figure}[!tb]
    \centering
    \includegraphics[width=0.95\textwidth]{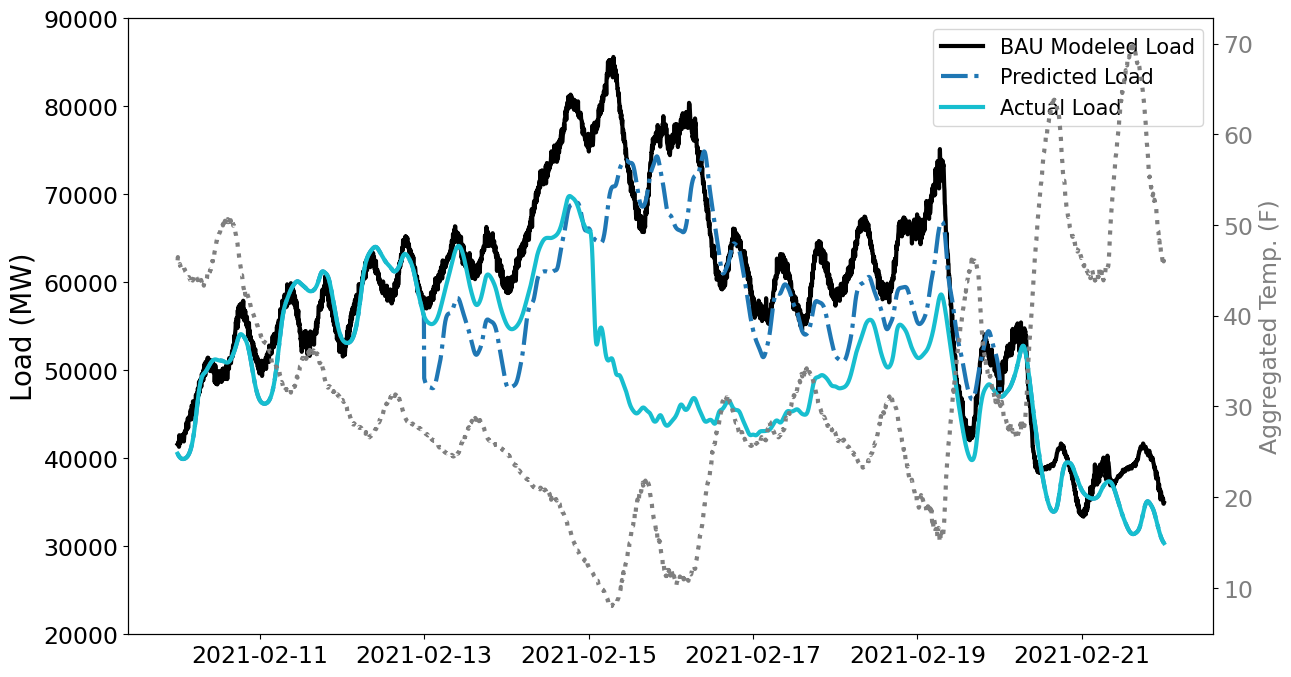}
    \caption{Simulated load for extreme condition and its comparison with actual (with outages) and predicted (without outages \cite{gruber2022profitability}) load.}
     \label{fig:time-series-comparison-extreme-condition}
\end{figure}

\subsection{BAU Modeled Case Under Extreme Conditions}\label{sec:bau-results}
\Cref{fig:time-series-comparison-extreme-condition} compares the simulated load profile with the actual (with outages) and predicted load data (without outages) for the Texas region, during the extreme winter temperatures in 2021.The actual load (with outages), i.e., the demand that was served by ERCOT during the extreme winter temperatures is taken from~\cite{ercot_hist}. For extreme weather prediction comparison purposes, data from \cite{gruber2022profitability} is used to represent a ``without outages'' condition. In \cite{gruber2022profitability} used a regression-based load model to predict ERCOT load during winter storm Uri if there had been no outages. The presented comparison of this paper with \cite{gruber2022profitability} successfully demonstrates that the presented modeling and simulation platform while modeling at-scale physics-based distribution grid loads, is also able to capture aggregated load modeling trends, as proposed in the state-of-the-art literature. From \cref{fig:time-series-comparison-extreme-condition}, it can be seen that the simulated load matches well with the predicted load from \cite{gruber2022profitability}. As the prediction model in \cite{gruber2022profitability} was only presented for the days when outages were experienced, both predicted and actual loads are the same on the non-outages days. The simulated load shows that as temperatures drop to historic low values, the demand increases to represent the increased heating demand. From the 13$^{\text{th}}$ of February onward, the ERCOT system started experiencing outages, which is not captured in the simulation model.
\subsection{Space Heating Electrification and Efficiency Measures Cases (Case 1, 1a, 1b and 1c)}
\begin{figure}[!ht]
\centering\includegraphics[width=0.95\textwidth]{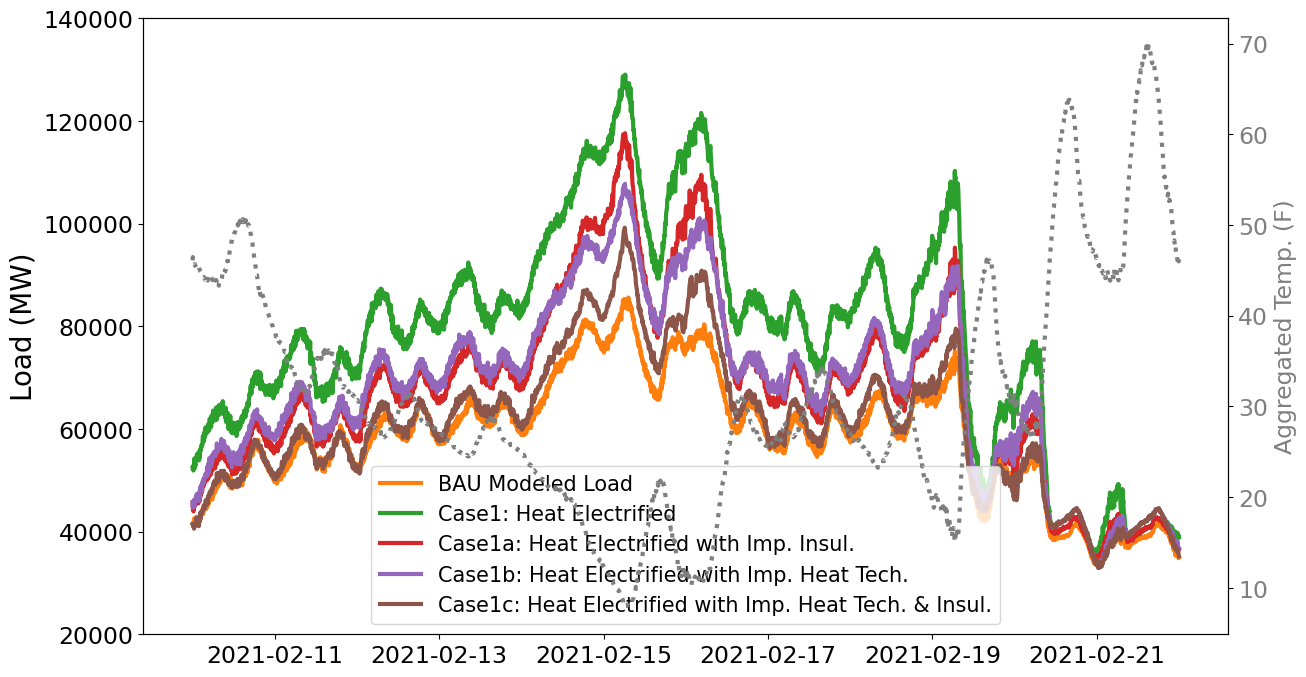}
    \caption{Simulated load profiles for the electrification scenarios and its comparison with BAU modeled case under extreme weather conditions.}
     \label{fig:time-series-comparison-electrification}
\end{figure}

\Cref{fig:time-series-comparison-electrification} shows simulated load time-series for extreme temperature days for all cases associated with space heating electrification and relevant measures and their comparison to BAU modeled case under extreme conditions of \cref{sec:bau-results}.

The results from electrification cases are summarized below:
\begin{itemize}
    \item Case 1, which considers all space heating electrified without any measures, models the highest peak load of $\sim$122 GW. This is an increase of 52.5\% as compared to the peak load obtained from BAU modeled case ($\sim$80 GW). All peak loads occur at the same time during 15$^{th}$ February.
    \item Measures of improving only insulation (Case 1a), heating system technology (Case 1b), and combination of these measures (Case 1c) predicts $\sim$118 GW, $\sim$108 GW and $\sim$99 GW of peak load, respectively. That is, as compared to Case 1, Cases 1a, 1b and 1c, decreases the peak load by 3.4\%, 12.9\%, and 18.8\%, respectively. Eventually, Case 1c, which can be considered as an efficient space heating case, models an increase of 23.7\%, as compared to BAU modeled load.

    \item As a comparison between higher insulation levels (Case 1a) versus better heating equipment (Case 1b), it can be seen that for extreme winter temperatures, better heating equipment (case 1b) reduces the electrified load levels more than higher insulation (case 1a). However, during usual winter temperatures (other than 13$^{\text{th}}$-20$^{\text{th}}$ February), both cases (case 1a and 1b) yield almost the same system loading. As an example how HVAC load and dynamics change due to the change in the parameters of houses, \cref{sec:app_house_model} provides an example of a house modeled in with (Case 1a) and without (Case 1) high insulation.
\end{itemize}
\begin{figure}[!h]
\centering    \includegraphics[width=0.95\textwidth]{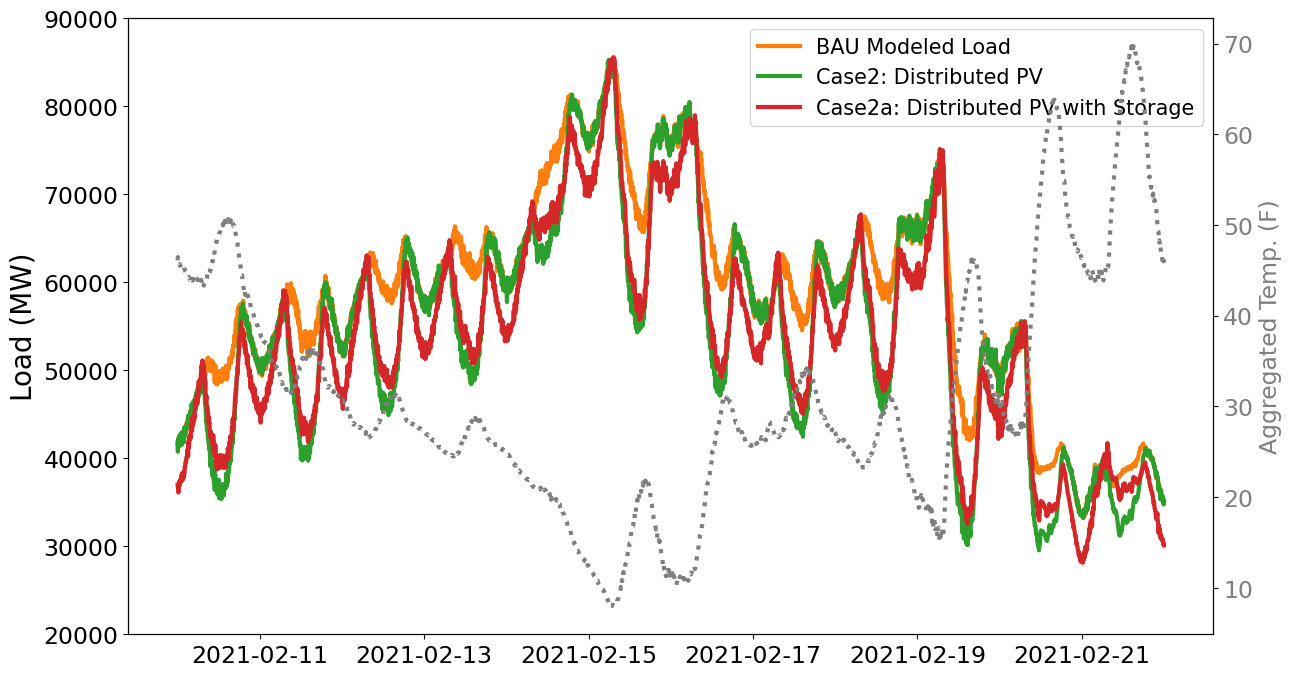}
    \caption{Simulated load profiles for distributed PV and storage integration adoption cases and its comparison to BAU Modeled load under extreme weather days.}
     \label{fig:time-series-comparison-renewables}
\end{figure}

\subsection{Distributed PV and Storage Integration Cases (Case 2 and 2a)}
\begin{figure}[!h]
    \centering
\includegraphics[width=0.95\textwidth]{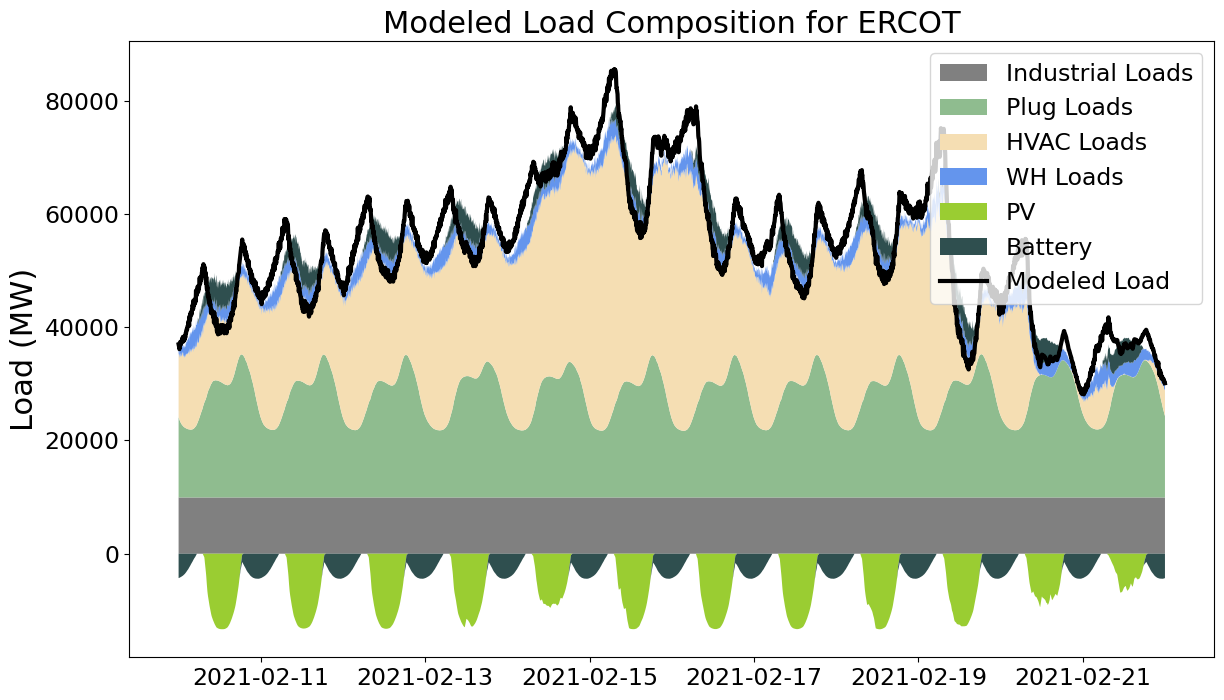}
    \caption{ERCOT load decomposition for the integration of distributed PV and storage (Case 2a).}
     \label{fig:DSO2-renewables}
\end{figure}

\Cref{fig:time-series-comparison-renewables} shows simulated load time-series for extreme temperature days for all renewable integration scenarios and their comparison with the BAU case. 
Summary of the results from the simulation of these cases are: 
\begin{itemize}
    \item Distributed PV reduces the load during day-time, due to solar PV production, however, as expected, it has no impact during evening and night-time peaks.
    \item Distributed storage helps to flatten the net load, by charging during the day-time and discharging during the evening-night-time. This has a positive impact during nights of the extreme temperature days, e.g., 15$^{\text{th}}$, 16$^{\text{th}}$ and 19$^{\text{th}}$ February. The highest peak load for BAU modeled case, distributed PV (Case 2) and distributed PV with storage (Case 2a), all occur around 5-6 am of 15$^{\text{th}}$ February, during these times there was neither considerable solar PV production nor distributed battery discharging energy available. To visualize this better, this interplay between the charging/discharging of distributed batteries and the solar PV production can be seen in \cref{fig:DSO2-renewables}, where to distinguish load and local generation support, aggregated solar PV production and discharging is shown as negative, and charging as positive values.
\end{itemize}

\begin{figure}[!h]
    \centering
    \includegraphics[width=0.95\textwidth]{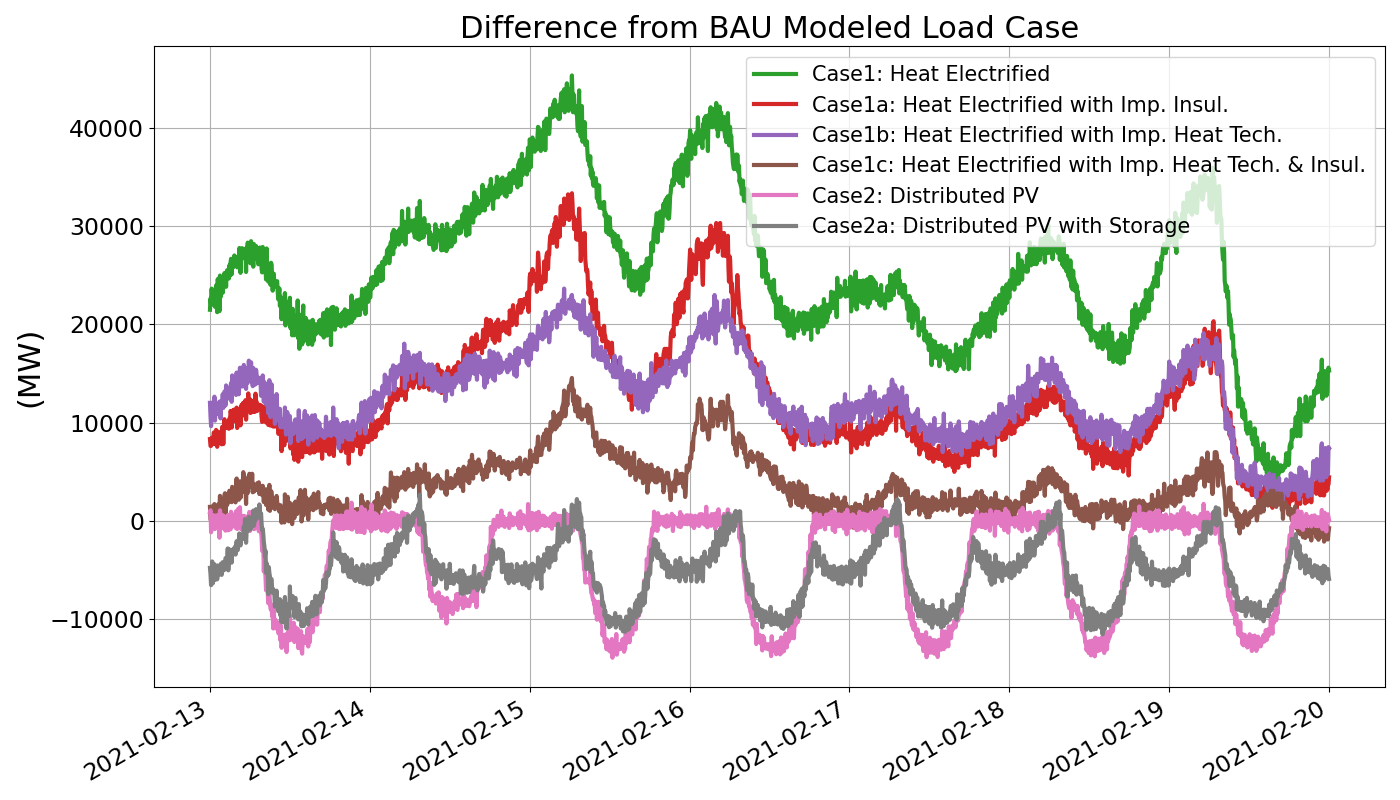}
    \caption{Load difference of alternative modeled cases with BAU modeled case for extreme weather days}
\label{fig:bau_load_diff}
\end{figure}
\begin{figure}[!h]
    \centering
    \includegraphics[width=0.95\textwidth]{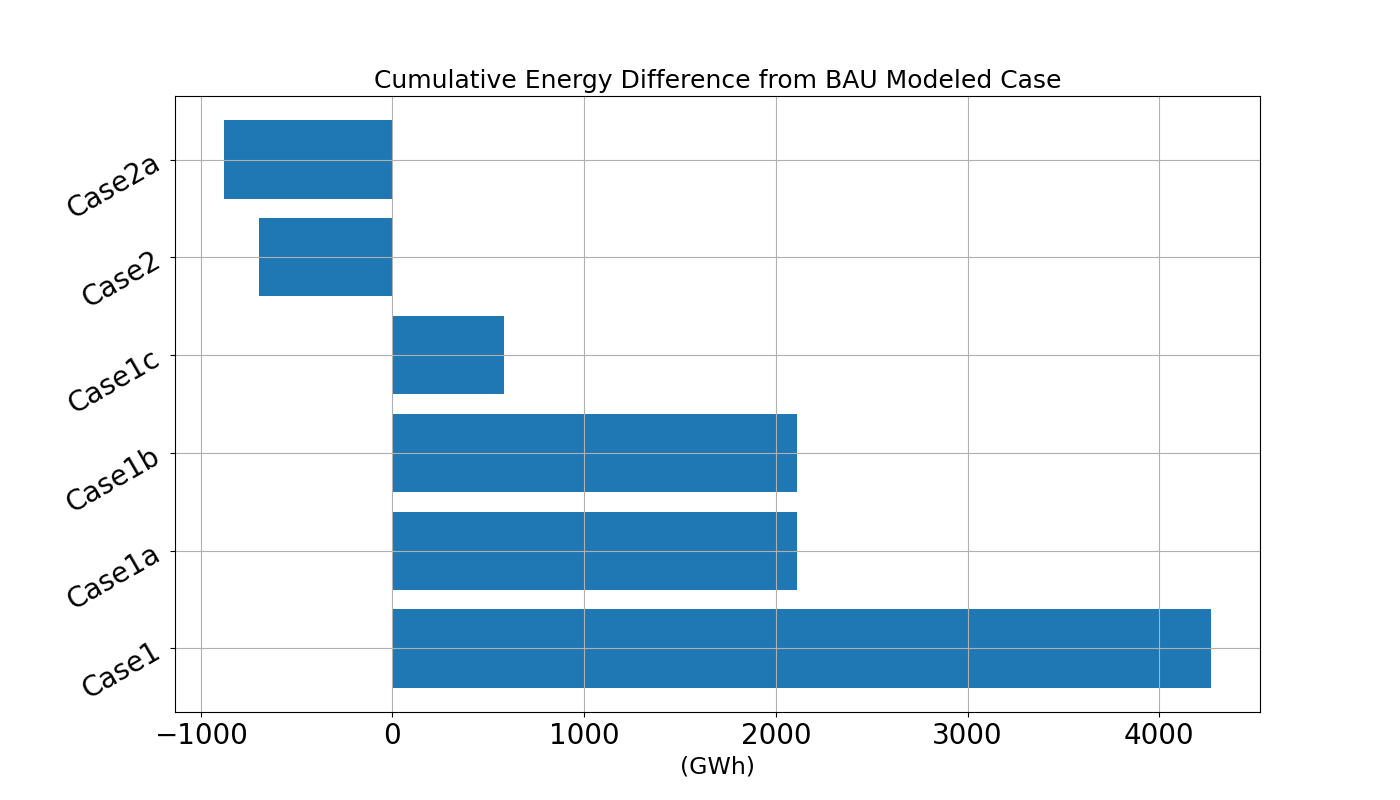}
    \caption{Energy difference of alternative modeled cases with BAU modeled case for extreme weather days}
     \label{fig:bau_en_diff}
\end{figure}
\begin{figure}[!h]
    \centering
\includegraphics[width=0.95\textwidth]{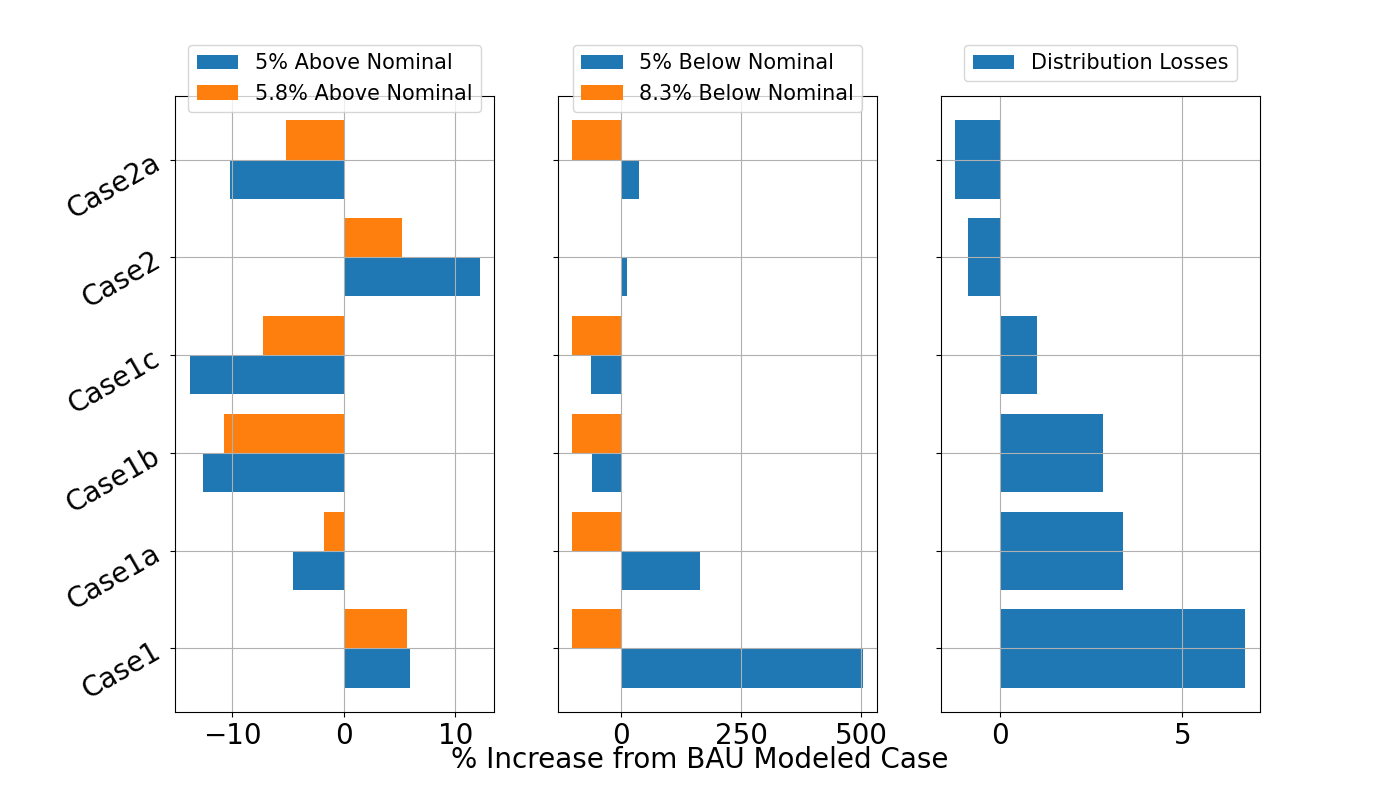}
    \caption{Comparison of grid violations and losses for alternative modeled cases with BAU modeled case.}
     \label{fig:grid_quant}
\end{figure}

\subsection{Case Comparisons Summary}
In this section, for extreme winter temperature days (13$^{\text{th}}$--20$^{\text{th}}$ Februaary), we compare all alternative cases on renewable integration and load electrification (Case 1, 1a, 1b, 1c, 2, 2a) against the 1) BAU modeled load case and 2) ERCOT supplied demand.
\subsubsection{Comparison Against BAU Modeled Case} As compared to BAU modeled case, \Cref{fig:bau_load_diff} and \cref{fig:bau_en_diff} show predicted demand and its corresponding energy difference of alternative future cases, respectively. Maximum ($\sim$45 GW) and minimum ($\sim$18 GW) load difference is observed for case 1 and Case 2, respectively. The inclusion of local generation in the form of distributed PV (Case 2) along with storage (Case 2a), reduces the total energy demand, as the net power imported from the upstream of the grid reduces. From \cref{fig:bau_en_diff}, it can be seen that Case 2a has a higher negative energy difference as compared to Case 2. This is because with the presence of storage (Case 2a), the higher utilization of PV and local load supply by batteries reduces grid losses and hence its overall demand.
\Cref{fig:grid_quant} shows in relative percentage, how alternative future scenarios changes grid violations and losses. The 5\% above/below and 5.8\%/8.3\% above/below voltage violations ranges are taken from standards voltage limits to be maintained and avoided, respectively~\cite{ansi-volt}. Due to large space heating electrification load, a very high percentage of low voltage counts occur (500\% increase from BAU modeled case). Interestingly, the reductions in the above-voltage violations count occur with improved heating technology (Case 1b) is more than with improve insulation (Case 1a) which is due to improved dynamics of the heating equipment. The inclusion of PV (Case 2), increases the above-voltage violation counts, as injecting local power causes rise in voltage. Comparing to BAU case, with the inclusion of storage (Case 2a), all voltage violations counts and losses are reduced.
\subsubsection{Comparison Against Actual ERCOT Supplied Demand During Winter Storm Uri}
\begin{figure}[!h]
    \centering
    \includegraphics[width=0.95\textwidth]{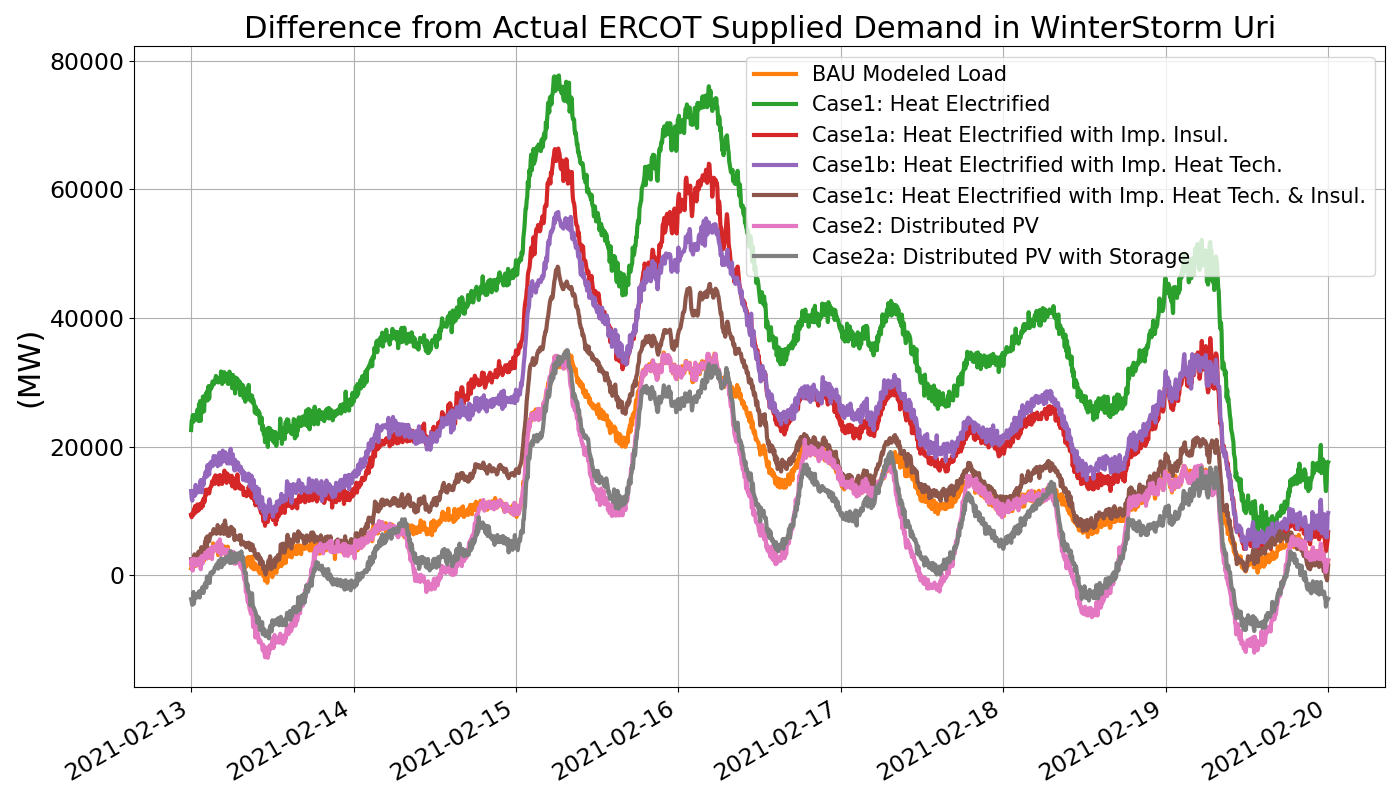}
    \caption{Load difference of alternative modeled cases with actual ERCOT supplied load for extreme weather days}
     \label{fig:ercot_load_diff}
\end{figure}

\begin{figure}[!h]
    \centering
\includegraphics[width=0.95\textwidth]{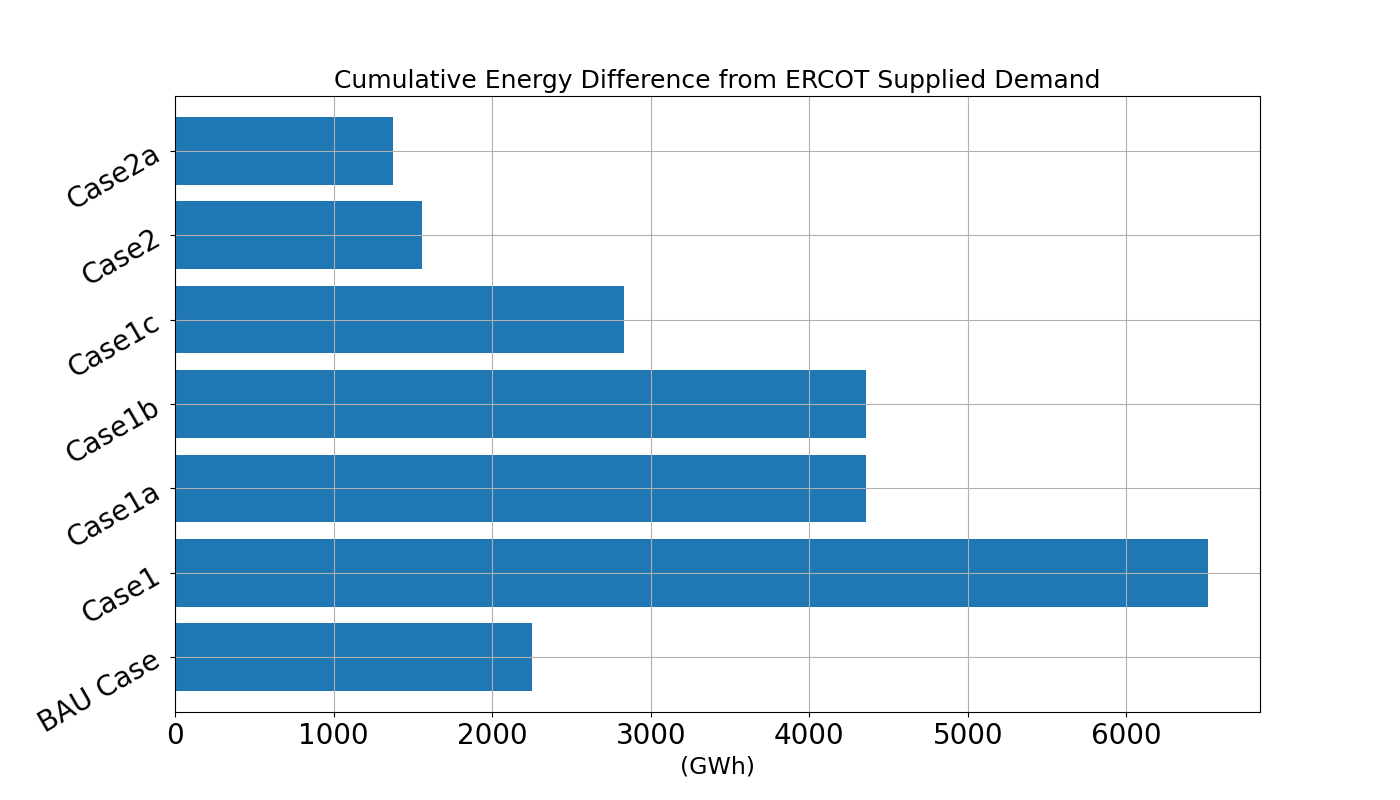}
    \caption{Energy difference of alternative modeled cases with actual ERCOT supplied load for extreme weather days}
     \label{fig:ercot_en_diff}
\end{figure}

In this subsection, we compare the difference of all alternative future cases against the actual demand that was supplied in ERCOT during the winter storm Uri. In doing so, we attempt to demonstrate how the presented analytical capability of this paper could be utilized to explore ``what-if'' scenarios of the required system flexibility, considering a retrospective extreme weather type. 
\Cref{fig:ercot_load_diff} and \cref{fig:ercot_en_diff} shows the difference of supplied ERCOT demand and its cumulative energy. BAU modeled load case demonstrates that $\sim$30 GW of maximum instantaneous load ($\sim$2.2 TWh energy) was potentially unmet during the winter storm, which could have jumped to $\sim$77 GW ($\sim$6.5 TWh energy ) for the electrified case (Case 1). For both Case 2 and Case 2a, the inclusion of local energy production and utilization bring the potential unmet energy to $\sim$1.5 TWh and $\sim$1.3 TWh, which is a decrease of $31\%$ and $40\%$, as compared to BAU modeled case, respectively. 

\section{Conclusion \& Future Work}\label{sec:conclusion}
The paper presented at-scale analysis of distribution grid system demand during extreme weather conditions. The proof-of-concept was demonstrated on winter storm Uri of February 2021. A modeling and simulation platform was presented which was shown to be extensible by constructing alternative future scenarios and their impact on the overall system demand during extreme conditions. The paper demonstrated that a physics-based at-scale modeling and simulation platform can realistically capture complex interactions between grid infrastructure, their operations procedures and the external influences. Hence, such a platform is advocated to investigate the resulting distribution system loads associated with future DER scenarios and mitigation schemes during extreme weather events. We demonstrate this in the paper by showing how electrification of the loads and the integration of renewable energy influence load dynamics and utilization of local generation on the grid infrastructure. Future works will be targeted towards the control of grid infrastructure, specially demand-side resources, to address challenges associated with grid resilience.
\section{Appendix}\label{sec:appendix}

\subsection{Modeling and Simulation Platform Implementation Overview}\label{sec:app_model}
In the current implementation of the presented platform \cref{sec:modelingframework}, the message bus consists of Hierarchical Engine for Large-scale Infrastructure Co-Simulation Hierarchical Engine for Large-scale Infrastructure Co-Simulation (HELICS)\cite{helics}, which serves as a bridge to exchange relevant information between these two simulators and controls the flow of the simulation at the desired time step (30-seconds for this paper). For the power flow simulator, we use GridLAB-D, due to its capability to model the weather-dependent load dynamics~\cite{chassin:2014aa}. For the weather simulator, we use time-indexed data files in a ``.csv'' format with the relevant information required by the power flow simulator. The results of the simulation platform are stored in a ``.h5'' file format and a post-processing script is written to visualize them. Depending upon the study case, (e.g. business-as-usual, extreme conditions, higher penetration of solar photo-voltaic (PV), etc.), a Python script was used to coordinate and set-up the simulation platform.

\begin{table}[t]
\centering
\caption{Region characteristics}
\label{Region characteristics}
 \begin{adjustbox}{max width=\textwidth}
\begin{tabular}{|c|c|c|c|c|}
\hline 
 Region number (Utility Type)& Modeled Feeders&Modeled Houses &Modeled Area&Scaling Factor\\
\hline
Region 1 (Urban)& \begin{tabular}{@{}c@{}}R4-12.47-1 \\ R4-12.47-2\end{tabular}	&893&	Dallas, TX&	3816.95\\
\hline
Region 2 (Urban)&\begin{tabular}{@{}c@{}}R5-12.47-1 \\ R5-12.47-2\end{tabular}	&1308	&Houston, TX&	2351.17\\
\hline
Region 3 (Rural)&R5-12.47-5	&1539	&Lamar, TX	&58.14\\
\hline
Region 4 (Rural)&R5-12.47-5	&1539	&Midland, TX	&479.44\\
\hline
Region 5 (Urban)&\begin{tabular}{@{}c@{}}R5-12.47-1 \\ R5-12.47-2\end{tabular}	&1308	&Hays, TX	&1395.95\\
\hline
Region 6 (Urban)&\begin{tabular}{@{}c@{}}R4-12.47-1 \\ R5-12.47-1\end{tabular}	&1525	&Val Verde, TX	&67.95\\
\hline
Region 7 (Suburban)&R5-12.47-5	&1539	&Nueces, TX	&895.00\\
\hline
Region 8 (Rural)&R5-12.47-5	&1539&	Presidio, TX&	23.55\\
\hline
\end{tabular}
\end{adjustbox}
\end{table}

\subsection{ERCOT Distribution System Modeling}\label{sec:app_Regiont_sim}
This section gives a brief overview of the modeled region. For a detailed explanation of the process, the interested readers are referred to \cite{reeve2022dso+}.

To capture the system demand of Texas due to extreme events, similar to \cite{reeve2022dso+}, multiple Regions were modeled and aggregated to represent the total system load. A summary of each modeled Region is presented in \cref{Region characteristics}. Each Region consists of multiple prototypical feeders as the backbone infrastructure (e.g., topology, rated equipment loading, power conversion elements, and power delivery elements) \cite{Schneider2008taxonomy}. Statistical data available for the region, number/types of customers (residential, customer, and industrial), and the peak load of the region. The "Feeder Generator" process shown in \cref{fig:modeling_simulation_platform} then distributes the population of the feeder with the houses, to match the rated load of the distributed houses with the statistically observed load region. With this process, the GridLAB-D model gets populated with both residential and commercial loads with their corresponding ETP model and its relevant parameters to reflect desired loading characteristics of the Region.  The parameters are extracted using building types, their construction year, the HVAC system installed and its efficiency, referred to as Coefficient of Performance (COP). 

Collectively, all Regions ended up modeling 11929 buildings (residential and commercial), among which 8952 HVAC units and 4923 water heaters were modeled. To manage the complexity and accuracy of the desired load response, a lower number of GridLAB-D houses were modeled than the recorded total number of customers in the modeled region. A scaling factor was then used to represent the modeled load at the Region level. For example, for the residential load of the Region, the final load profile's scaling factor is calculated as: Total Residential Customer in the Region/(Residential Customers Fraction Among Total Customers $\times$ Number of Residential Homes Modeled in GridLAB-D). As aggregated residential and commercial load is then obtained, a constant industrial load (due to lack of data) is then added to obtain the total Region load.

\subsection{GridLAB-D ETP Model Overview}\label{sec:app_etp}
GridLAB-D models two types of loads. One with the control loop and the other without the control loop.  ZIP load objects (constant impedance, current, and power) are modeled to model the end-use loads such as lights and plugs. Thermostatic loads (HVAC and water heaters) are modeled with the control loop. With the combination of the HVAC and ZIP model, the total load is modeled, and the corrected voltage response is provided to capture the impact of load on the power flows. One of the largest loads in the house is the HVAC load and as it is sensitive to outdoor temperature and helps the simulation platform to capture the impact of extreme weather conditions, a small overview of the model is given next. 

\begin{figure}[!h]
    \centering
    \includegraphics[width=0.95\textwidth]{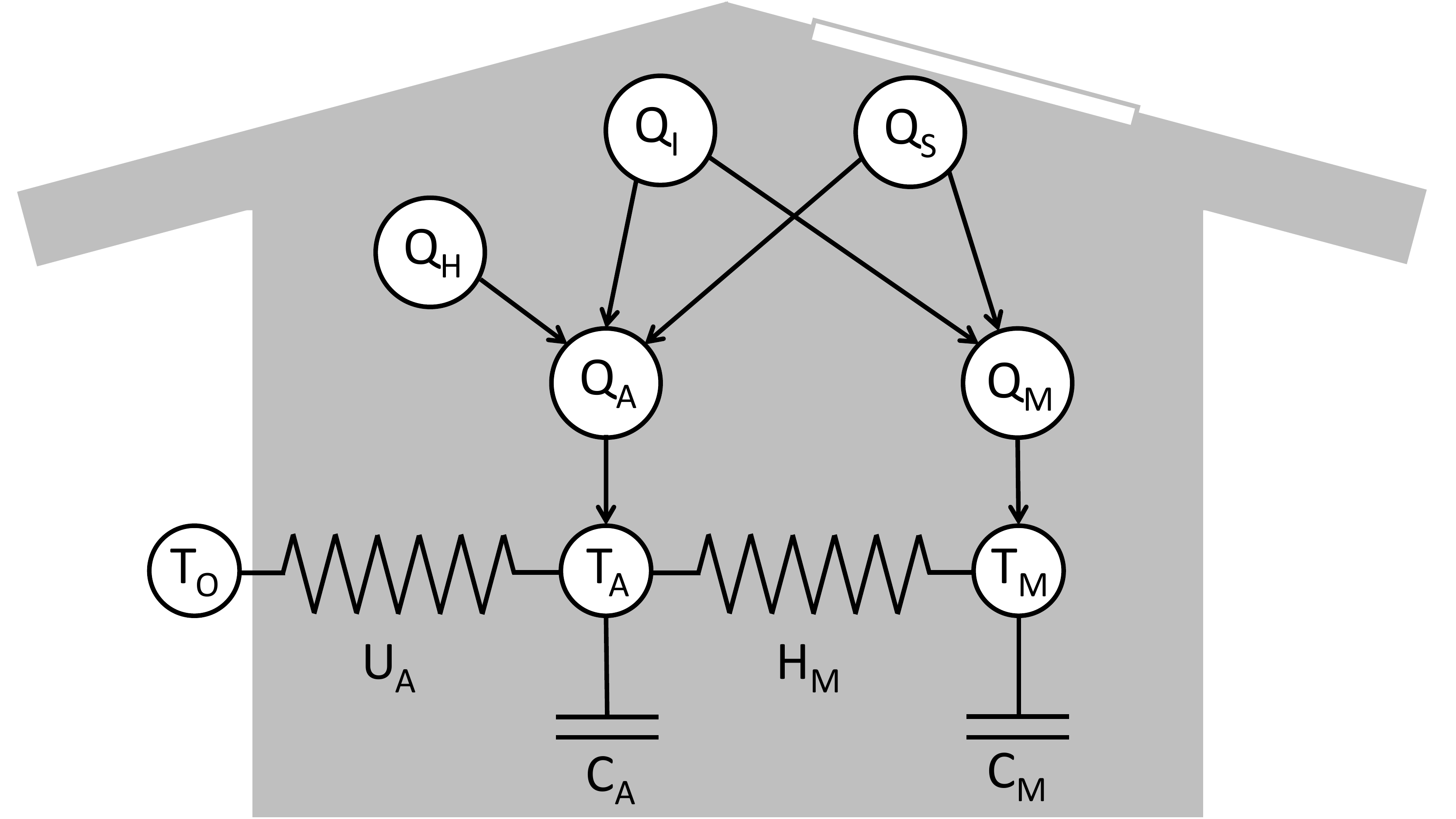}
    \caption{ETP model of a house in GridLAB-D, refer to \cref{sec:app_etp} for more information on ETP model development.}
     \label{fig:etp_model}
\end{figure}

GridLAB-D models the HVAC system using the equivalent thermal parameter (ETP) approach. This captures the essential response of the house under various circumstances such as weather, occupant behavior, appliances, heating, ventilation, and air-conditioning (HVAC) system operation to analyze the grid operation~\cite{GLD}. Fig. \ref{fig:etp_model} illustrates the ETP model of a house in GridLAB-D and the following equations explain how dynamics of loads are captured:
\begin{align}
& Q_A – U_A(T_A – T_O) – H_M (T_A – T_M) – C_A \frac{d}{dt}T_A = 0 \label{eq:Tabalance} \\
& Q_A = Q_H + (1- f_{i}) Q_I + (1-f_{s}) Q_S \label{Qa}\\
& U_A = A_gU_g + A_d/R_d + A_w/R_w + A_c/R_c \nonumber \\
& \quad + A_f/R_f + 0.018 ACH \label{Ua} \\
&  \frac{d}{dt} T_A = \frac{1}{C_A} \left(  Q_A – U_A(T_A – T_O) – H_M (T_A – T_M)  \right) \label{dTa} 
\end{align}

Overall, the electrical power consumed by the HVAC system is calculated by the system rated power and fan power. In addition, it considers motor losses that are related to the efficiency of the induction motor when the electric cooling system or heat pump for heating is utilized.


\makenomenclature
\nomenclature{\(Q_A\)}{Total heat gain to the indoor air}
\nomenclature{\(Q_H\)}{Internal heat gain by HVAC operation}
\nomenclature{\(Q_I\)}{Internal heat gain by non-HVAC equipment}
\nomenclature{\(Q_S\)}{Solar radiation}
\nomenclature{\(U_A\)}{Total heat loss coefficient (conductance)}
\nomenclature{\(H_M\)}{Internal mass surface conductance}
\nomenclature{\(C_A\)}{Total air mass}
\nomenclature{\(T_A\)}{Indoor air temperature}
\nomenclature{\(T_O\)}{Outdoor air temperature}
\nomenclature{\(T_M\)}{Building mass temperature}
\nomenclature{\(f_s\)}{Solar gain fraction to mass}
\nomenclature{\(f_i\)}{Internal gain fraction to mass}
\nomenclature{\(R_d\)}{R-value, doors}
\nomenclature{\(R_c\)}{R-value, ceilings}
\nomenclature{\(R_w\)}{R-value, walls}
\nomenclature{\(R_f\)}{R-value, floors}
\nomenclature{\(U_g\)}{U-value, windows}
\nomenclature{\(A_d\)}{Area, doors}
\nomenclature{\(A_c\)}{Area, ceilings}
\nomenclature{\(A_w\)}{Area, walls}
\nomenclature{\(A_f\)}{Area, floors}
\nomenclature{\(A_g\)}{Area, windows}
\nomenclature{\(ACH\)}{Air changes per hour}

To provide a better quality of indoor circumstances, HVAC systems operate to cool or heat the buildings. The ETP model expresses the physical characteristics of the house with a state-space control form. The heat balance (conservation of energy) for the air temperature node $T_A$ is represented as \cref{eq:Tabalance}.
It is determined by how much thermal energy is stored in the air and mass of the building and how much heat can gain or lost from outside based on the actual physical properties of the building. 
For example, internal mass surface conductance $H_m$ is the total heat transfer coefficient by the building surface (exterior walls, interior walls, ceilings).
Total heat gain to the indoor air $Q_A$ is estimated by the heat gain or loss from non-HVAC equipment $Q_I$, solar radiation $Q_S$, and the HVAC operation $Q_H$ as shown in \cref{Qa}.
Total heat loss coefficient $U_A$ is the sum of all heat loss coefficients through the envelope of the building (walls, windows, doors, ceilings, floors, and infiltration air flows). 
Eventually, the indoor air temperature ($T_A$) changes can be estimated by heat gain or loss through the envelope of the building, weather conditions, internal heat gain, and HVAC operation as shown in \cref{dTa}.
\subsection{Impact of Insulation on HVAC load and Indoor Temperature Dynamics of a Building}\label{sec:app_house_model}
\begin{figure}[h]
    \centering   \includegraphics[width=0.95\textwidth]{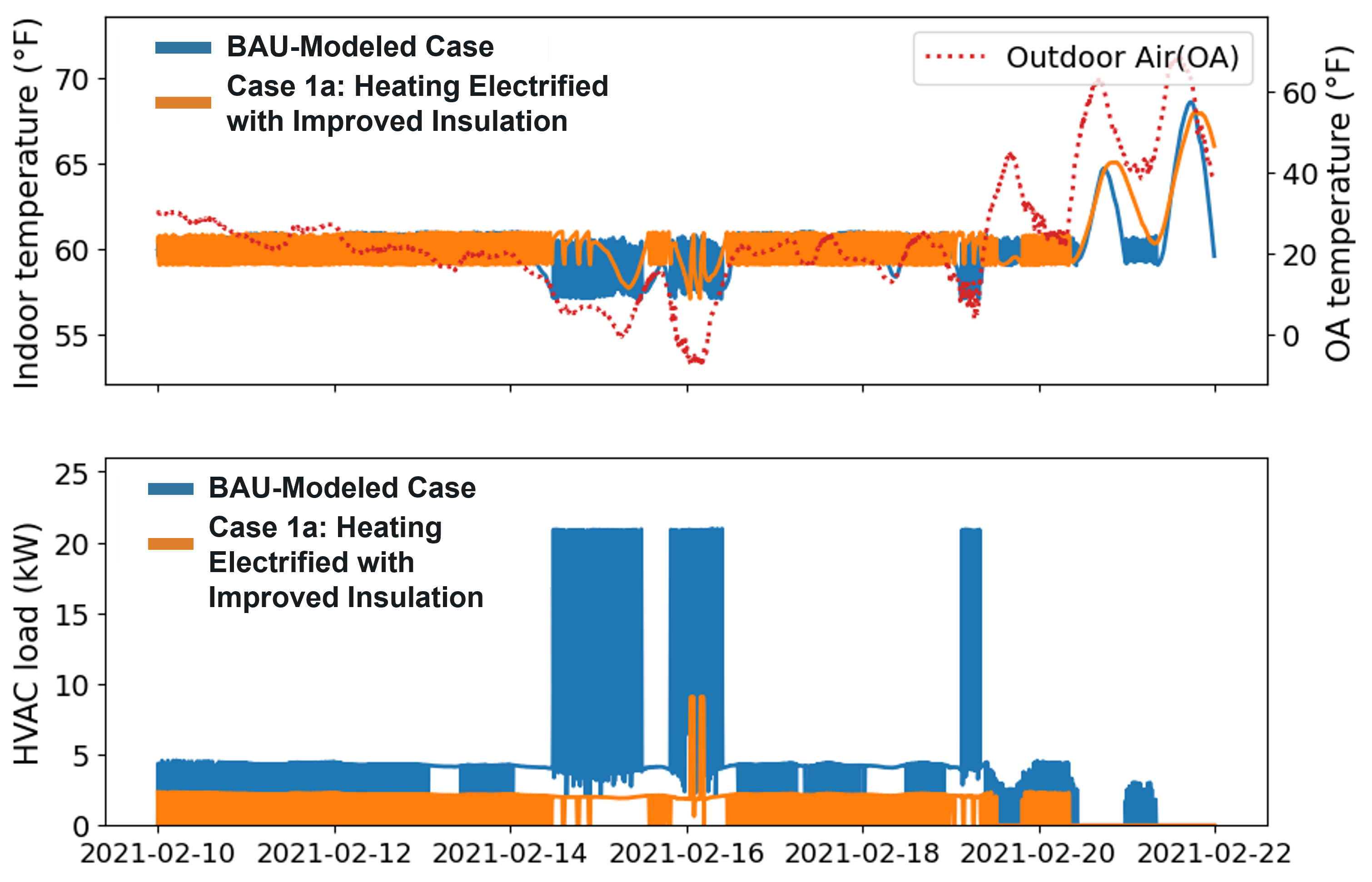}
    \caption{Indoor air temperature (top subplot) and HVAC load (bottom subplot) for comparison between non-insulated (from Case 1) and high-insulated building (from Case 1a).}     \label{fig:house_comparison_temp}
\end{figure}
\Cref{fig:house_comparison_temp} shows the indoor air temperature (top subplot) and HVAC loads (bottom subplot) for comparison between non-insulated building (Case 1) and high-insulated building (Case 1a). It highlights how the insulation affects the indoor air temperature and HVAC loads in the same building. Both buildings are heated with a heat pump with electric resistance auxiliary backup. The indoor air temperature in both cases was maintained in the desired temperature range by operating the heating systems, except for the extreme outdoor air temperature times (around 2/15 to 2/16). Also, at the extreme outdoor air temperature duration, both cases provided heating with the heat pump and the auxiliary system together, and it resulted in a high HVAC load. When it got warmer from 2/21 to 2/22, both buildings didn't require heating during the daytime but the non-insulated building (Case 1) operated heating during the night. The highly insulated building (Case 1a) maximized the benefits of thermal mass by reducing the heat flow between the indoor space and ambient so it preserved the daytime indoor temperature longer from outdoor air temperature fluctuations. Eventually, the non-insulated building (Case 1) required more heating operation. In addition, the HVAC system size in the non-insulated building (Case 1) is even larger than the highly insulated building (Case 1a) since it requires more heating/cooling, thus it resulted in a huge difference in HVAC loads between the cases.

\printnomenclature

 \bibliographystyle{elsarticle-num} 
 \bibliography{cas-refs}
\end{document}